 \documentclass[mnsc,sglanonrev]{informs4_hide}
\RequirePackage{tgtermes}
\RequirePackage{newtxtext}
\RequirePackage{newtxmath}
\usepackage{mathtools} 
\RequirePackage{bm}
\RequirePackage{endnotes}

\OneAndAHalfSpacedXII 

\usepackage{algorithm}
\usepackage{algpseudocode}
\usepackage{tikz}

\usepackage{natbib}
 \bibpunct[, ]{(}{)}{,}{a}{}{,}%
 \def\bibfont{\small}%
\usepackage{anyfontsize}
\usepackage{enumitem}
\usepackage[colorlinks,
            linkcolor=blue,
            anchorcolor=blue,
            citecolor=blue,
            urlcolor=blue,
            hypertexnames=false
            ]{hyperref}
            
\usepackage{subcaption}
\usepackage{autonum}

\usepackage{forloop}
\newcommand{\defcal}[1]{\expandafter\newcommand\csname
c#1\endcsname{{\mathcal{#1}}}}
\newcommand{\defbb}[1]{\expandafter\newcommand\csname
bb#1\endcsname{{\mathbb{#1}}}}
\newcommand{\defbf}[1]{\expandafter\newcommand\csname
bf#1\endcsname{{\mathbf{#1}}}}
\newcounter{calBbCounter}
\forLoop{1}{26}{calBbCounter}{
	\edef\letter{\Alph{calBbCounter}}
	\expandafter\defcal\letter
	\expandafter\defbb\letter
	\expandafter\defbf\letter
}

\newcommand{\rnum}[1]{\lowercase\expandafter{\romannumeral #1\relax}}
\newcommand{\Rnum}[1]{\uppercase\expandafter{\romannumeral #1\relax}}

\newcommand{\ai}{\mathsf{AI}}
\newcommand{\doc}{\mathsf{DM}}
\newcommand{\atten}{\mathsf{Atten}}
\newcommand{\comp}{\mathsf{Comp}}

\renewcommand{\qed}{\null\nobreak\hfill\ensuremath{\blacksquare}}

\EquationsNumberedThrough    

\TheoremsNumberedThrough     
\ECRepeatTheorems  %

\MANUSCRIPTNO{MS-BST-2024-08471.R2}

\begin{document}


\RUNAUTHOR{Li et al.}

\RUNTITLE{Attribution and Persuasion: The Paradox of Interpretable AI}

\TITLE{Attribution and Persuasion: The Paradox of Interpretable AI}

\ARTICLEAUTHORS{%
\AUTHOR{Hanzhe Li}
\AFF{University of Hong Kong, \EMAIL{hanzhe.li@connect.hku.hk}}

\AUTHOR{Jin Li}
\AFF{University of Hong Kong, \EMAIL{jli1@hku.hk}}

\AUTHOR{Ye Luo}
\AFF{University of Hong Kong, \EMAIL{kurtluo@hku.hk}}

\AUTHOR{Xiaowei Zhang}
\AFF{Hong Kong University of Science and Technology, \EMAIL{xiaoweiz@ust.hk}}
} 

\ABSTRACT{%
This paper studies AI persuasion by distinguishing between two reasons for disagreement: attention differences, where the AI detects features the decision-maker missed, and comprehension differences, where the AI and the decision-maker interpret observed features differently. We show that AI is more effective in persuading the decision-maker when the disagreement is due to attention differences rather than comprehension differences. We also show that the AI’s interpretability shapes how the decision-maker attributes the sources of disagreement and, in turn, whether they follow the AI’s recommendation. Our main result is that making AI uninterpretable can actually enhance persuasion and, in the presence of career concerns, improve decision accuracy.
}%





\maketitle


\newcommand{\chapquote}[3]{\begin{quotation} \textit{#1} \end{quotation} \begin{flushright} - #2, \textit{#3}\end{flushright} }

\section{Introduction}\label{sec intro}

\chapquote{People should stop training radiologists now. It's just completely obvious within five years deep learning is going to do better than radiologists.}{Geoffrey Hinton}{2016}

\chapquote{The big claims about AI assume that if something is possible in theory, then it will happen in practice. That is a big leap.}{Peter Cappelli and Valery Yakubovich}{2024}

Artificial Intelligence (AI) holds the potential to transform decision-making across the economy, from financial analysts forecasting market trends and HR managers screening applicants to judges determining bail and doctors diagnosing diseases~\citep{agrawal2018prediction,brynjolfsson2021,agrawal2022power}. Yet, despite the promise of superior predictive performance, AI adoption in high-stakes managerial environments remains slower than expected~\citep{Dranove2022,Slack2024,D-HKU2025,mcelheran2025rise}. A recurring barrier is the reluctance of human experts to incorporate algorithmic suggestions into their final decisions—a phenomenon often termed ``algorithm aversion'' (e.g., \citealt{dietvorst2015}, \citealt{Longoni2019}, \citealt{Kawaguchi_2020}, and \citealt{liu2023algorithm}).

While there are many reasons for algorithm aversion, a critical but under-explored factor is the \emph{career concern} of the decision-maker (DM). In many managerial contexts, the decision to rely on AI is not merely a choice about accuracy; it is a signal about the DM's own ability. This creates a distinct talent assessment problem specific to the age of AI. In the era of traditional Information and Communication Technologies (ICT), utilizing a tool (like a spreadsheet or a database) was widely viewed as an enhancement to worker productivity; better performance indicated a more effective worker. However, with AI, this conclusion can reverse. Because AI performs the core cognitive task of judgment, a manager who systematically defers to an algorithm when their opinions conflict may signal that their own expertise is inferior or redundant. Consequently, a DM concerned with their professional image may reject valid AI advice to demonstrate their own competence, leading to suboptimal organizational outcomes \citep{Arkes2007,Victoria2013,almog2025ai}.

In this paper, we examine how the design of AI, specifically its interpretability, affects the DM's willingness to resolve disagreements with the algorithm. We propose that disagreements between a human expert and an AI arise from two distinct sources: \emph{attention differences} and \emph{comprehension differences}.

An attention difference occurs when the AI identifies specific data points or features that the DM overlooked. For instance, an AI might flag a clause in a legal contract that a lawyer missed, or an anomaly in an X-ray that a radiologist failed to see. In these cases, the AI provides objective evidence of an oversight. Correcting this mistake improves accuracy without necessarily invalidating the DM's core judgment capabilities; the AI acts as an observational aid.

A comprehension difference arises when both the DM and the AI observe the same features but assign them different weights or interpretations. For example, a hiring manager and an AI might both notice a candidate's gap in employment; the manager interprets it as a red flag, while the AI calculates it has no correlation with future performance. Similarly, a financial analyst might diverge from a model's stock pick not because of missing data, but because they disagree on the predictive weight of recent interest rate hikes. In these scenarios, the disagreement is a matter of judgment. Here, changing their mind requires the DM to concede that the AI possesses superior comprehension. This type of disagreement places the DM and the AI in competition where the human-AI interaction is turned into a comparison of who has better comprehension skill.

We highlight these two distinct sources of disagreement because they differ in their effectiveness in persuading the DM. If the disagreement is due to an attention difference (AI detects an anomaly that the DM overlooks), then AI always persuades the DM by pointing it out. By contrast, if the disagreement is because of differences in comprehension, AI persuades the DM only when the DM has sufficiently worse comprehension skill.

Our analysis focuses on how the DM attributes the source of a disagreement when the AI's reasoning for its recommendation may not be known. We define an AI as interpretable if the DM can see not only the AI's recommendation but also the features it uses. Conversely, an uninterpretable AI reveals only the final recommendation.

Our main result is that making AI uninterpretable can actually enhance persuasion and, in the presence of career concerns, improve decision accuracy. To see why this is the case, note that when the AI is uninterpretable, the DM carries out \emph{Bayesian attribution} and assigns positive probability to an attention difference. Because the attention difference is more persuasive, a weighted average of comprehension and attention differences can persuade the DM, even if the comprehension difference alone will not. Making AI uninterpretable, therefore, enhances persuasion by allowing the more persuasive attention difference to ``subsidize'' the less persuasive comprehension difference. We refer to this mechanism as the \emph{averaging effect}. 

The weight the DM assigns to the attention difference depends on her skill level. A DM with lower attention skills (who is more likely to overlook positive features) is then more likely to attribute the disagreement to the attention difference. The higher the weight put on the attention difference, the more likely the DM is to change her opinion. Consequently, making AI uninterpretable is particularly effective in improving persuasion when the DM has lower attention skills. We refer to this mechanism as the \emph{attribution effect}.

We further show that this design choice has implications for career concerns. Consider a setting where DMs vary in their comprehension skills. High-skilled DMs naturally rely on their own comprehension. Low-skilled DMs, wishing to be perceived as high-skilled, will mimic this behavior by ignoring the AI, even when the AI is correct. This mimicry leads to inaccurate decisions.

By making the AI uninterpretable, we can alleviate the career-concern problem of the low-skilled DMs. Because uninterpretability increases the persuasiveness of the AI (via the averaging effect), high-skilled DMs become more willing to listen to it. This, in turn, reduces the reputational stigma of relying on the algorithm, allowing low-skilled DMs to follow the AI without revealing their type, thereby improving aggregate decision quality.

While we use medical diagnosis as our primary motivating application to illustrate the formal model (as features naturally correspond to symptoms or image abnormalities), our framework applies broadly to the managerial contexts introduced above: from hiring decisions and legal strategy to financial forecasting. In all these domains, the visibility of the ``why'' behind the AI's decision determines whether the human expert views the interaction as valid support or a challenge to their status. The paper sheds light on human-AI interaction and information design by showing that ``black box'' algorithms, often criticized for a lack of transparency, may possess a hidden advantage: they mitigate the career incentives that drive humans to reject algorithmic advice.

Finally, we show that, beyond career concerns, uninterpretability can also improve decision-making by strengthening the incentive to acquire information. When the DM disagrees with the AI, she can sometimes collect more evidence by re-reading a contract clause, ordering a confirmatory test, or consulting a colleague. We show that uninterpretability can make the DM more willing to collect such information. When the AI discloses the inputs behind its recommendation, the DM can assess whether the recommendation seems sensible without checking those inputs herself, which encourages her to lean on the AI. When the AI provides only its recommendation, the DM cannot resolve the disagreement by inspecting the AI's cited evidence, so additional evidence becomes more valuable at the margin. As a result, uninterpretability can induce more investigation and sometimes improve decision accuracy even though the AI is less transparent.\medskip

\noindent\textbf{Related literature.} 
Our paper is related to four broad literatures. First, a growing literature investigates how AI complements workers, often examining how AI affects workers of different skills; see, for example, \citet{Gruber2020}, \citet{Allen2022}, \citet{brynjolfsson2023}, \citet{Noy2023}, \citet{Jia2024}, and \citet{Wang2024}.\footnote{See \citet{ide2024} for the theoretical implications on the organization of knowledge work.} In these papers, the heterogeneity of workers is measured in a single dimension.  Our paper highlights the multidimensional skills of DMs. We show that the effectiveness of AI assistance depends not just on the overall skills of the DM, but also on the level of skills the DM has on each dimension: the DM benefits more when she has low attention skills than when she has low comprehension skills.

Second, our paper contributes to the literature on AI aversion (e.g., \citealp{dietvorst2015, Longoni2019, Luo2019, Kawaguchi_2020, Kim2024, Yin2024}; see \citealp{burton2020}, \citealp{Mahmud_2022}, and \citealp{de2023psychological} for surveys). \citet{Tong2021} document that employees, once they know that the feedback is from an AI, will lower their trust in the feedback and more concern themselves with the risks of job replacement. \citet{liu2023algorithm} find that drivers are less likely to follow AI's recommendations if they contradict past experiences and peers' actions. In the context of doctor-AI interaction, \citet{Jussupow2021} demonstrate that AI advice may undermine doctors' causal reasoning. \citet{Kang2024} find similar effects among financial analysts. \citet{Lebovitz2022} report that doctors tend to explain disagreements away when AI recommendations conflict with their own judgments. \citet{Agarwal2023} document deviations from rational updating among doctors. We contribute to this literature by showing how uninterpretability can make AI recommendations easier to accept and thus increase AI adoption.

Third, our paper is related to the theoretical literature on human-AI interaction. This literature examines how human-AI interactions are affected by various underlying factors and design choices. \citet{Agrawal_2019, NBERc14010} focus on different skills between AI and doctors. \citet{athey2020} examine the optimal allocation of decision rights between agents and AI. \citet{dai2020} analyze how career concerns affect the acquisition of information by doctors. We add to this literature by studying how the interpretability of AI affects the way the DM attributes the source of the disagreement.

Finally, this paper is related to the literature on information design; see \citet{Bergemann2019} and \citet{Kamenica2019} for general reviews of how information provision affects players' behavior. Our paper contributes to this literature by studying AI as an information provider and examining its interpretability design. In the context of reputation concerns, several studies show that non-transparency can improve welfare by disciplining reputational incentives \citep{Prat2005,levy2007decision,ashworth2010does,fox2012costly,FU201415,DEMORAGAS2022104282,Li2025}. More broadly, reputation concerns can distort organizational decision-making and have nontrivial effects on performance \citep{Fu2022}. We extend this line of inquiry by examining transparency design in the presence of explicit disagreements. Our results highlight the benefits of two types of nontransparency: uninterpretability, which hides why the AI disagrees with the DM, and private use of AI, which hides whether the DM deferred to the AI when they disagree.

\section{Model}\label{sec Model}
\smallskip
\subsection{Latent Fundamentals for Decision: State, Features, and Critical Dimension}

A DM needs to infer the underlying \emph{state} of an object to make a choice. The state is not directly observed. Instead, it is encoded in an underlying feature in one of two dimensions. This dimension is \emph{critical}: if its feature were observed, the state would be known exactly. The other dimension is \emph{non-critical}: its feature is pure noise and carries no information about the state. The two features and the critical dimension are also not directly observed. We refer to the state, the two features, and the critical dimension as the latent \emph{fundamentals} of the decision problem.

Formally, let $Z$ denote the underlying state and $\varepsilon$ denote noise. There are two dimensions, ``left'' ($L$) and ``right'' ($R$), with corresponding features $X(L)$ and $X(R)$. Let $C\in\{L,R\}$ denote the critical dimension, which is unknown to the DM. We assume that the feature in the critical dimension equals $Z$, while the other feature equals $\varepsilon$:
\begin{equation} \label{eq DGP}
\left\{\begin{aligned}
X(L)=&\bbI(C=L)Z+\bbI(C\neq L)\varepsilon, \\
X(R)=&\bbI(C=R)Z+\bbI(C\neq R)\varepsilon,
\end{aligned}
\right.
\end{equation}
where $\bbI$ is the indicator function. For example, if $C=L$, then $X(L)=Z$ and $X(R)=\varepsilon$. 

For simplicity, assume that $Z$ and $\varepsilon$ are binary, taking values in $\{0,1\}$, so $X(L)$ and $X(R)$ are binary as well. Following common terminology in machine learning, we call the state or a feature ``positive'' if it equals $1$ and ``negative'' if it equals $0$. The exact values of the decision fundamentals $(Z,X,C)$ are unknown and follow certain probability distributions.  
\begin{assumption}\label{assump:state}
\begin{enumerate}[label=(\roman*), noitemsep,topsep=0pt]
    \item The state is positive with probability $\gamma\in(0,1)$: $\Pr(Z=1)=\gamma$.
    \item The noise is independent of $Z$ and is positive with probability $\lambda\in(0,1)$: $\Pr(\varepsilon=1)=\lambda$.
    \item The two dimensions are equally likely to be critical, independent of $(Z,\varepsilon)$: $\Pr(C=L)=\frac{1}{2}$.
\end{enumerate}
\end{assumption}

\subsection{Signals for Decisions: Attention and Comprehension}
The DM relies on two signals for her decision: an \emph{attention} signal and a \emph{comprehension} signal, which are noisy versions of $X$ and $C$, respectively.
The attention signal represents the DM's observation of the two features $(X(L), X(R))$, while the comprehension signal reflects her judgment about which dimension is critical. Let $X^\doc \coloneqq (X^\doc(L),X^\doc(R))$ denote the DM's attention signal, where $X^\doc(j)\in\{0,1\}$ for both $j\in \{L, R\}$, and let $C^\doc\in\{L, R\}$ denote her comprehension signal.

In addition to her own signals, the DM has access to an AI that provides informational assistance. The AI receives its own pair of signals: an attention signal $X^{\ai} \coloneqq (X^{\ai}(L),X^{\ai}(R))$ and a comprehension signal $C^\ai$. 
Our modeling of AI as an attention--comprehension pair is motivated by the prevalent use of artificial neural networks in AI technologies: $X^{\ai}$ represents a vector of input data, and $C^\ai$ reflects how AI weights different input dimensions.\footnote{In medical applications, the AI's attention signal corresponds to the abnormalities that the AI flags in medical images.} Conditional on the fundamentals $(Z,C,\varepsilon)$, the signals $X^\doc(L)$, $X^\doc(R)$,
$C^\doc$, $X^\ai(L)$, $X^\ai(R)$, and $C^\ai$ are mutually independent. In this sense, the AI and the DM have separate observations about the features $X$ and separate comprehension regarding the critical dimension $C$.

We call the AI \emph{interpretable} if its signals are known to the DM. In contrast, the AI is \emph{uninterpretable} if the DM cannot observe its signals, but only its recommendation (which is derived from the AI's signals and will be elaborated later). We prefer the terms ``interpretable'' and ``uninterpretable'' over ``transparent'' and ``nontransparent'', because an AI can be uninterpretable even if it is partially transparent. For example, consider a radiologist diagnosing breast cancer by examining X-ray images. Even if a transparent AI flags several abnormalities, the radiologist may still find it uninterpretable as she does not know how the AI assigns importance to these abnormalities.

Moreover, we refer to the values of $C^\doc$ and $C^\ai$ as the \emph{DM's critical dimension} and the \emph{AI's critical dimension}, respectively. We make the following assumptions regarding the DM and AI's signals.

\begin{assumption}\label{assumption: signals}
    \begin{enumerate}[label=(\roman*), noitemsep,topsep=0pt]
        \item Neither the DM nor the AI makes Type \Rnum{1} (i.e., false positive) errors in observing features:  $\Pr(X^i(j)=0\mid X(j)=0)=1$ for $i\in\{\doc, \ai\}$ and $j\in\{L,R\}$.\footnote{This assumption represents an extreme case in which observing a positive feature is conclusive evidence. However, we can also allow nonzero Type~\Rnum{1} and Type~\Rnum{2} errors without affecting our results. This alternative assumption is discussed in Remark~\ref{remark on Proposition 1} of Section~\ref{sec Main}, and Online Appendix~\ref{Appendix signal errors} presents related results formally.}
        \item Both the DM and the AI may make Type \Rnum{2} (i.e., false negative) errors in observing features: 
        $\pi^i\coloneqq \Pr(X^i(j)=1\mid X(j)=1) \in [0, 1]$ 
        for $i\in\{\doc, \ai\}$ and $j\in\{L, R\}$.
        \item Both the DM and the AI may miscomprehend the critical dimension: 
        $p^i \coloneqq \Pr(C^i = C\mid C)\in[\frac{1}{2},1]$ for $i\in\{\doc, \ai\}$.
    \end{enumerate}
\end{assumption}

\subsection{Decision Process}
The DM's decision process with the AI's assistance is as follows:
\begin{enumerate}[noitemsep,topsep=0pt]
    \item Nature determines the state $Z$, noise $\varepsilon$, and the critical dimension indicated by $C$. The feature $X$ is then generated according to \eqref{eq DGP} and the value of $C$.
    \item The DM receives signals  $(X^\doc, C^\doc)$ and then makes an \emph{initial} decision $D\in\{0,1\}$.
    \item The AI receives signals $(X^{\ai},C^\ai)$ and then makes a recommendation $A\in\{0,1\}$.
    \item The DM observes 
    \begin{itemize}[noitemsep,topsep=0pt]
        \item $(X^{\ai},C^\ai)$ if the AI is interpretable, or 
        \item $A$ if the AI is uninterpretable.
    \end{itemize} 
    The DM then updates her belief about the state $Z$ to make the \emph{final} decision $F\in\{0,1\}$.
\end{enumerate} 
Each decision or recommendation (i.e., $D$, $A$, or $F$) is determined by the posterior probability of the state being positive given the corresponding information set. If this probability is greater than or equal to $\frac{1}{2}$, then the decision or recommendation is assigned a value of $1$ (``positive''). Otherwise, it is assigned a value of $0$ (``negative'').\footnote{Formally, the rational DM has utility $u(F,Z)=\bbI(F=Z)$ and chooses $F$ to maximize her expected utility $\mathbb{E}[u(F,Z)\mid \mathcal{I}]=\Pr(F=Z\mid \mathcal{I})$ given her information set $\mathcal{I}$. Since $F\in\{0,1\}$, this is equivalent to choosing $F=1$ if and only if $\Pr(Z=1\mid \mathcal{I})\geq\frac{1}{2}$, and the same criterion applies to the initial decision $D$ and the AI's recommendation $A$. Such thresholding parallels classification tasks performed by AI technologies in fields such as computer vision and natural language processing, where the threshold is not necessarily $\frac{1}{2}$ but is typically adjusted between $0$ and $1$ for each task to balance the costs of false positive and false negative errors \citep{Goodfellow-et-al-2016}. For simplicity, we assume the threshold is $\frac{1}{2}$ in our model.}  Throughout the paper, we focus on a rational DM who is risk-neutral and uses Bayes' rule to update her belief, and discuss extending our model to behavioral DMs only at the end of Section~\ref{sec career}.

The quality of the DM's initial decision (i.e., without the AI's assistance) is fully determined by the accuracy of the pair of signals $(X^\doc, C^\doc)$ relative to the underlying $(X, C)$. Accordingly, we refer to $(\pi^\doc, p^\doc)$ as the DM's \emph{attention skill} and \emph{comprehension skill}. Likewise, $(\pi^\ai, p^\ai)$ are termed the AI's attention skill and comprehension skill.

The decision process in our model reflects settings where the DM must combine information from multiple dimensions. For example, when diagnosing breast cancer from mammography, a doctor looks for abnormalities in the images and assesses how likely each finding is to indicate malignancy \citep{Lebovitz2022}. In promotion decisions, a manager reviews several performance measures and evaluates a candidate's readiness by weighing how predictive each measure is for success in the new role \citep{Grabner2013}. Our model allows us to study how AI affects the DM's willingness to revise the initial decision when it conflicts with the AI's recommendation. By highlighting that disagreement can arise from multiple sources, we show that the reason for disagreement matters.

\begin{remark}\label{remark on complementarity}
In our model, attention and comprehension capture distinct aspects of information acquisition: attention governs which features are observed, while comprehension governs how observed features are interpreted. As a result, these AI skills benefit the DM through different channels. In Online Appendix~\ref{appendix AI complementarity}, we study how AI skills interact with the DM's comprehension skill in a manner similar to \citet{BORGERS2013165}. While their paper asks whether one signal increases the value of another, we ask how AI attention and AI comprehension change the accuracy gain from improving the DM's comprehension. We show that AI skills complement the DM's comprehension when disagreement stems from attention differences and substitute for it when disagreement arises from comprehension differences.
\end{remark}

\subsection{Preliminaries}
To simplify our analysis, we make the following assumptions regarding the values of $\gamma$, $\lambda$,  $(\pi^\doc, p^\doc)$, and $(\pi^\ai, p^\ai)$. Recall that $\gamma$ denotes the probability that the state is positive, and $\lambda$ denotes the probability that the noise is positive.

\begin{assumption}\label{assumption gamma lambda 1}
    $\lambda\leq\gamma<\frac{1}{2}$.
\end{assumption}

\begin{assumption}\label{assumption gamma lambda 2}
    $(1/\gamma+1/\lambda-2)p^i+\pi^i>1/\lambda$ for $i\in\{\doc, \ai\}$.
\end{assumption}

Assumption~\ref{assumption gamma lambda 1} states that the state is positive with probability less than $\frac{1}{2}$. Additionally, the probability that the noise is positive is even smaller. This implies that, under our data generating process \eqref{eq DGP}, the feature in the critical dimension is more likely to be positive than the feature in the non-critical dimension. Assumption~\ref{assumption gamma lambda 2} ensures that when the DM or the AI observes a positive feature only in the non-critical dimension, they will believe that the state is less likely to be positive than negative.\footnote{Formally, this can be shown through the following likelihood ratio: \[\frac{\Pr[Z=1\mid X^i(C^i)=0,X^i(-C^i)=1]}{\Pr[Z=0\mid X^i(C^i)=0,X^i(-C^i)=1]}=\frac{\gamma}{1-\gamma}\frac{\Pr[X^i(C^i)=0,X^i(-C^i)=1\mid Z=1]}{\Pr[X^i(C^i)=0,X^i(-C^i)=1\mid Z=0]}<1,\] where $i\in\{\doc,\ai\}$ and $\{C^i\}\cup\{-C^i\}=\{L,R\}$.} Therefore, in this case, neither the DM nor the AI will make a positive decision or recommendation.

We maintain Assumptions~\ref{assump:state}--\ref{assumption gamma lambda 2} throughout the paper. Under these assumptions, the DM will make a positive initial decision if and only if the DM observes a positive feature in her critical dimension. An analogous statement applies to the AI's recommendation. Lemma~\ref{lem benchmark} states this formally.

\begin{lemma}[\textbf{Decision Through 
Critical Dimension}]\label{lem benchmark}
    The DM makes a positive initial decision (i.e., $D=1$) if and only if $X^\doc(C^\doc)=1$. 
    The AI makes a positive recommendation (i.e.,  $A=1$) if and only if $X^\ai(C^\ai)=1$.
\end{lemma}

\section{AI Persuasion}\label{sec Main}

Will the DM be persuaded to change her initial decision when it conflicts with the AI's recommendation? How does the AI's interpretability affect persuasiveness? This section examines these questions by focusing on the case in which the DM initially makes a negative decision ($D=0$) and then receives a positive recommendation from the AI ($A=1$). We then briefly discuss the opposite case, which yields similar insights.

\subsection{Interpretable AI}

When the AI is interpretable, the DM can observe the AI's attention and comprehension signals $(X^\ai, C^\ai)$. 
This allows the DM to identify the source of disagreement, which may arise either from an \emph{attention difference} or a \emph{comprehension difference}.

\begin{figure}[htbp]
    \FIGURE{
    \includegraphics[width=0.7\textwidth]{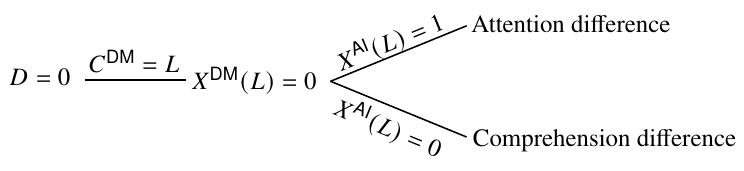}}
    {Two Sources of Disagreement When the DM's Initial Decision Is Negative ($D=0$) and the AI Recommends a Positive Decision ($A=1$).
    \label{fig: two sources}}
    {}
\end{figure}

Assume the DM has initially made a negative decision. Lemma~\ref{lem benchmark} indicates that she must have observed a negative feature in her critical dimension, $X^\doc(C^\doc)=0$, while the AI's positive recommendation implies $X^\ai(C^\ai)=1$. We distinguish two sources of disagreement according to what the AI observes in the DM's critical dimension.

\begin{definition}[\textbf{Sources of Disagreement}]\label{def: disagreement}
Suppose $D=0$ and $A=1$. The disagreement is
\begin{enumerate}[label=(\roman*),noitemsep,topsep=0pt]
    \item an \emph{attention difference} if the AI observes a positive feature in the DM's critical dimension, i.e., $X^\ai(C^\doc)=1$; and
    \item a \emph{comprehension difference} if the AI observes a negative feature in the DM's critical dimension, i.e., $X^\ai(C^\doc)=0$.
\end{enumerate}
\end{definition}

Under an attention difference, the AI observes a positive feature in the DM's critical dimension that the DM missed. Under a comprehension difference, both the DM and the AI observe the same negative feature there ($X^\ai(C^\doc)=X^\doc(C^\doc)=0$), yet the AI still recommends a positive decision; by Lemma~\ref{lem benchmark}, its recommendation is then necessarily driven by a different critical dimension in which it observes a positive feature ($C^\ai\neq C^\doc$ and $X^\ai(C^\ai)=1$).

Definition~\ref{def: disagreement} organizes disagreement by what the AI reveals in the DM's critical dimension, rather than by which of the AI's signals differ. To see that the two differences are exhaustive, note that disagreement can be described by two binary distinctions: whether the comprehension signals differ ($C^\ai$ versus $C^\doc$), and whether the AI observes a positive feature in the DM's critical dimension ($X^\ai(C^\doc)$). Of the four combinations, the case $C^\ai=C^\doc$ with $X^\ai(C^\doc)=0$ cannot arise under $A=1$, since by Lemma~\ref{lem benchmark} the AI would then recommend $A=0$. The remaining three are:
\begin{enumerate}[label=(\roman*),noitemsep,topsep=2pt]
    \item the AI looks at the very same dimension as the DM and observes a positive feature there: $C^\ai=C^\doc$ and $X^\ai(C^\doc)=1$;
    \item the AI looks at a different dimension and observes no positive feature in the DM's: $C^\ai\neq C^\doc$ and $X^\ai(C^\doc)=0$; and
    \item the AI looks at a different dimension yet still observes a positive feature in the DM's: $C^\ai\neq C^\doc$ and $X^\ai(C^\doc)=1$.
\end{enumerate}
Our attention difference comprises (\rnum{1}) and (\rnum{3}), and our comprehension difference is (\rnum{2}). Since $X^\ai(C^\doc)$ is binary, the two are mutually exclusive and exhaustive. We group (\rnum{1}) and (\rnum{3}) together because, as we show next, both persuade the DM for the same reason: the AI reveals a positive feature in her critical dimension. We now examine how persuasion varies with the source of disagreement and the DM's skill.

\begin{lemma}[\textbf{Interpretable AI: Known Attribution}]\label{lemma I interpretable AI}
    Suppose that the AI is interpretable and disagreement is given by $D=0$ and $A=1$. Then, the following holds:
    \begin{enumerate}[label=(\roman*),noitemsep,topsep=0pt]
        \item \label{part:interp-att} When an attention difference occurs, the AI persuades the DM to change her decision from $D=0$ to $F=1$ regardless of her skill $(\pi^\doc,p^\doc)$.
        \item \label{part:interp-comp} When a comprehension difference occurs, the AI persuades the DM if and only if her comprehension skill $p^\doc$ is weakly below a threshold $p_I$. The threshold is increasing in the AI's comprehension skill $p^\ai$.
\end{enumerate}
\end{lemma}

\begin{figure}[htbp]
\FIGURE{
\begin{subfigure}{0.4\textwidth}
\includegraphics[width=\linewidth]{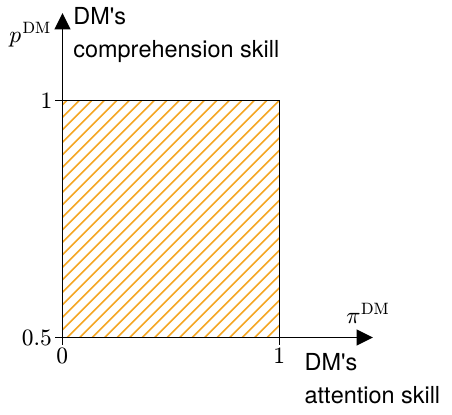} 
\caption{Attention persuasion}
\label{fig: attention persuasion}
\end{subfigure}
\hspace{0.05\textwidth}
\begin{subfigure}{0.4\textwidth}
\centering
\includegraphics[width=\linewidth]{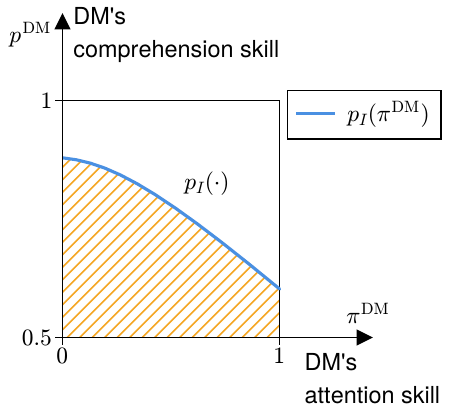}
\caption{Comprehension persuasion}
\label{fig: comprehension persuasion}
\end{subfigure}
}
{The Range of DMs Persuaded by the Interpretable AI.
\label{fig: persuasion with known attribution}}
{}
\end{figure}

Part~\ref{part:interp-att} of Lemma~\ref{lemma I interpretable AI} shows that when an attention difference occurs, the AI successfully persuades the DM regardless of her skill level. This unconditional persuasion occurs because the attention difference provides clear evidence of the DM's oversight: she missed a positive feature in her critical dimension that the AI correctly identified. This corresponds to our assumption that attention signals induce no Type~\Rnum{1} error. By providing a positive signal in the DM's critical dimension, the AI enables her to correct her attentional mistake and revise her initial negative decision.

In contrast, Part~\ref{part:interp-comp} of Lemma~\ref{lemma I interpretable AI} indicates that, under a comprehension difference, the AI can persuade the DM only if the DM's comprehension skill $p^\doc$ falls below a specific threshold $p_I$. In this case, the disagreement arises from different judgments regarding which dimension is critical for decision making. Unlike the attention difference, the comprehension difference does not provide conclusive evidence that the DM has made a mistake. Because comprehension signals can contain both Type~\Rnum{1} and Type~\Rnum{2} errors, the DM cannot definitively confirm that her original judgment was incorrect. Persuasion therefore hinges on the DM's assessment of her own comprehension: she accepts the AI's recommendation only when she views her comprehension as sufficiently unreliable, that is, when $p^\doc \le p_I$. Because a more skilled AI is more trustworthy in judging which dimension is critical, the threshold $p_I$ increases with the AI's comprehension skill $p^\ai$.

A key insight from our analysis of interpretable AI is that persuasion depends on why the DM and the AI disagree. Across tasks, disagreement can come from missed evidence or from different interpretations of the same evidence. As a result, AI can be persuasive in some settings but resisted in others. For example, \citet{Lebovitz2022} find that radiologists rely on AI more in lung-cancer diagnosis than in breast imaging. Lung cancer diagnosis depends mostly on attention to spotting the nodule. When disagreements arise because of overlooked evidence, it is easier for the doctors to change their opinion. In contrast, breast imaging depends more on the interpretation of the anomaly. This means that disagreements more often reflect comprehension differences and are harder for doctors to accept.

The same logic extends beyond medicine. In hiring, AI should be more persuasive when it points to overlooked facts, such as missing credentials, inconsistencies, or relevant experience, and less persuasive when it primarily reweights signals the manager has already considered. In financial consulting, AI should be most persuasive when it flags information the analyst missed and less persuasive when it offers a different take on the same market information.\smallskip

\begin{remark}\label{remark on Lemma 2}
Assumption~\ref{assumption: signals} rules out Type~\Rnum{1} errors in attention signals, and we use this simplifying feature to build intuition for Lemma~\ref{lemma I interpretable AI}. This assumption, however, is not essential. In Online Appendix~\ref{Appendix signal errors}, we allow both Type~\Rnum{1} and Type~\Rnum{2} errors and derive a threshold for the Type~\Rnum{1} error rate such that, when it lies below this threshold and the Type~\Rnum{2} error rate remains positive, the conclusions of Lemma~\ref{lemma I interpretable AI} continue to hold.

Moreover, if we reverse the error structure so that attention signals never incur Type~\Rnum{2} errors but may incur Type~\Rnum{1} errors, the result in Lemma~\ref{lemma I interpretable AI} reverses as well. In that case, under disagreement $D=1$ and $A=0$, the DM's positive decision is overturned when an interpretable AI reveals a negative feature in her critical dimension. We discuss this type of disagreement further in Remark~\ref{remark pos to neg}.
\end{remark}

\subsection{Uninterpretable AI}
When the AI is uninterpretable, the DM observes only the AI's recommendation and not its observation of the features. Her information set is therefore $\mathcal{I}^{\doc} = (X^{\doc}, C^\doc, D=0, A=1)$. If she disagrees with the AI's recommendation, she carries out \emph{Bayesian attribution}: she uses Bayes' rule to infer whether the AI observed a positive feature in her critical dimension, that is, whether the disagreement is an attention difference or a comprehension difference.

Formally, we define the posterior probability of an attention difference as 
\[\Pr(\atten\mid \cI^\doc):=\Pr(X^\ai(C^\doc)=1\mid \cI^\doc),\] 
and the posterior probability of a comprehension difference as
\[\Pr(\comp\mid \cI^\doc):=\Pr(X^\ai(C^\doc)=0\mid \cI^\doc).\] 
Since $X^\ai(C^\doc)$ is binary, these two events partition the disagreement, and the DM can write her posterior belief that the state is positive as a weighted average over the two sources of disagreement:
\begin{equation}\label{eq: expression}
\begin{aligned}
\Pr(Z=1\mid \cI^\doc) =& \Pr(\atten\mid \cI^\doc) \cdot \Pr(Z=1\mid \atten, \cI^\doc) \\
&+ \Pr(\comp\mid \cI^\doc) \cdot \Pr(Z=1\mid \comp, \cI^\doc).
\end{aligned} 
\end{equation}

Since the DM aims to maximize decision accuracy, she is persuaded to change her initial decision if and only if
\begin{equation}\label{eq: decision criterion}
    \Pr(Z=1\mid \mathcal{I}^\doc) \geq \frac{1}{2}.
\end{equation}
The corollary below establishes basic properties of the Bayesian attribution decomposition in \eqref{eq: expression}, which we use to study when an uninterpretable AI persuades the DM.

\begin{corollary}[\textbf{Attention Difference Is More Persuasive}]\label{corollary 1}
    The AI is more persuasive when disagreement stems from an attention difference rather than a comprehension difference: $\Pr(Z=1\mid \atten, \cI^\doc)>\frac{1}{2}$ for any $(\pi^\doc,p^\doc)$, but $\Pr(Z=1\mid \comp, \cI^\doc)\geq\frac{1}{2}$ only if $p^\doc\leq p_I$.
\end{corollary}

To interpret Corollary~\ref{corollary 1}, suppose that the DM could observe the source of disagreement even though the AI is uninterpretable. Then she would update just as she would with an interpretable AI. The corollary therefore follows from Lemma~\ref{lemma I interpretable AI}.

Part~\ref{part:interp-att} of Lemma~\ref{lemma I interpretable AI} shows that an attention difference is always persuasive, regardless of the DM's skill. Hence, conditional on an attention difference, we have $\Pr(Z=1\mid \atten, \cI^\doc)\geq\frac{1}{2}$. Moreover, since with positive probability an attention difference reveals positive features in both the left and the right dimensions, the inequality is strict. By contrast, Part~\ref{part:interp-comp} shows that a comprehension difference is persuasive only when the DM's comprehension skill is below $p_I$. Equivalently, $\Pr(Z=1\mid \comp, \cI^\doc)\geq \frac{1}{2}$ holds only if $p^\doc\leq p_I$.

When the AI is uninterpretable, the DM cannot tell whether disagreement comes from attention differences or comprehension differences. Corollary~\ref{corollary 1} then implies that persuasion depends on the probability the DM assigns to the attention difference in her Bayesian attribution. In \eqref{eq: expression}, this probability is the weight $\Pr(\atten\mid \cI^\doc)$. Since the attention difference is always persuasive, the uninterpretable AI is persuasive whenever $\Pr(\atten\mid \cI^\doc)$ is large enough. The next proposition formalizes this condition.

\begin{proposition}[\textbf{Uninterpretability Enhances Persuasion}]\label{prop I enhance}
    Suppose that the AI is uninterpretable and disagreement is given by $D=0$ and $A=1$.
    Then, the following holds:
    \begin{enumerate}[label=(\roman*),noitemsep,topsep=0pt]
    \item \label{part:uninterp-thres} \textbf{\emph{Threshold for persuasion:}} There
      exists a threshold $p_U\le 1$ such that the AI persuades the DM whenever
      $p^\doc\leq p_U$.
    \item \label{part:uninterp-avg} \textbf{\emph{Averaging effect:}} $p_U\geq p_I$, and
      the inequality is strict if $\pi^\doc<1$ and $p^\ai<1$.
    \item \label{part:uninterp-att} \textbf{\emph{Attribution effect:}} On the range where
      $p_U<1$, the gap $p_U-p_I$ increases as $\pi^\doc$ decreases.
    \end{enumerate}
\end{proposition}

\begin{figure}[htbp]
    \FIGURE{
    \includegraphics[width=0.4\textwidth]{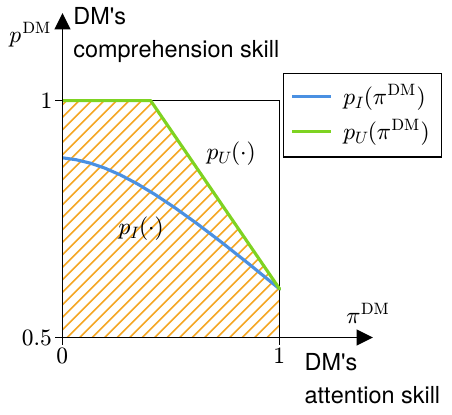}}
    {The Range of DMs Persuaded by the Uninterpretable AI. \label{fig: uninterpretability enhances persuasion}}
    {}
\end{figure}

Part~(i) of Proposition~\ref{prop I enhance} shows that the uninterpretable AI persuades a DM whenever the DM's comprehension skill is below a threshold $p_U$. Part~(ii) further shows that $p_U \geq p_I$, as illustrated in Figure~\ref{fig: uninterpretability enhances persuasion}. As a result, a DM with $p^\doc \in (p_I, p_U]$ may resist persuasion from an interpretable AI when there are comprehension differences, but always follows an uninterpretable one.

This result arises from what we call the averaging effect. With an interpretable AI, the DM updates differently depending on whether disagreement comes from an attention difference or a comprehension difference, which gives the two persuasion regions in Figures~\ref{fig: attention persuasion} and \ref{fig: comprehension persuasion}. With an uninterpretable AI, she cannot condition on the source of disagreement, so the persuasion region in Figure~\ref{fig: uninterpretability enhances persuasion} lies between those two benchmark regions. Formally, uninterpretability forces the DM to pool attention and comprehension differences and to form a weighted average of posteriors (see \eqref{eq: expression}). Since Corollary~\ref{corollary 1} shows that attention differences are more persuasive than comprehension differences, pooling shifts the posterior toward the attention-difference case. When $p^\doc$ is just above $p_I$, a comprehension difference alone yields $\Pr(Z=1\mid \comp, \cI^\doc) < \frac{1}{2}$, while an attention difference yields $\Pr(Z=1\mid \atten, \cI^\doc) > \frac{1}{2}$. The weighted average can therefore exceed $\frac{1}{2}$, so the DM is persuaded even though she would resist an interpretable AI under a comprehension difference.

In addition, Part~\ref{part:uninterp-att} of Proposition~\ref{prop I enhance} shows an \emph{attribution effect}. When the DM is less attentive, she is more likely to attribute the disagreement to an attention difference. This increases the weight on the attention difference in \eqref{eq: expression} and widens the gap between $p_U$ and $p_I$.

For intuition, suppose that the DM observes a positive feature only in her non-critical dimension. If her attention skill is lower, then conditional on the AI making a positive recommendation, it becomes more likely that she missed a positive feature in her critical dimension. She therefore assigns a higher probability to an attention difference. Since attention differences are more persuasive than comprehension differences, this shift makes the DM more likely to be persuaded. Thus, reducing interpretability is especially effective at enhancing persuasion when the DM is less attentive.

\begin{figure}[htbp]
    \FIGURE{\includegraphics[width=0.7\textwidth]{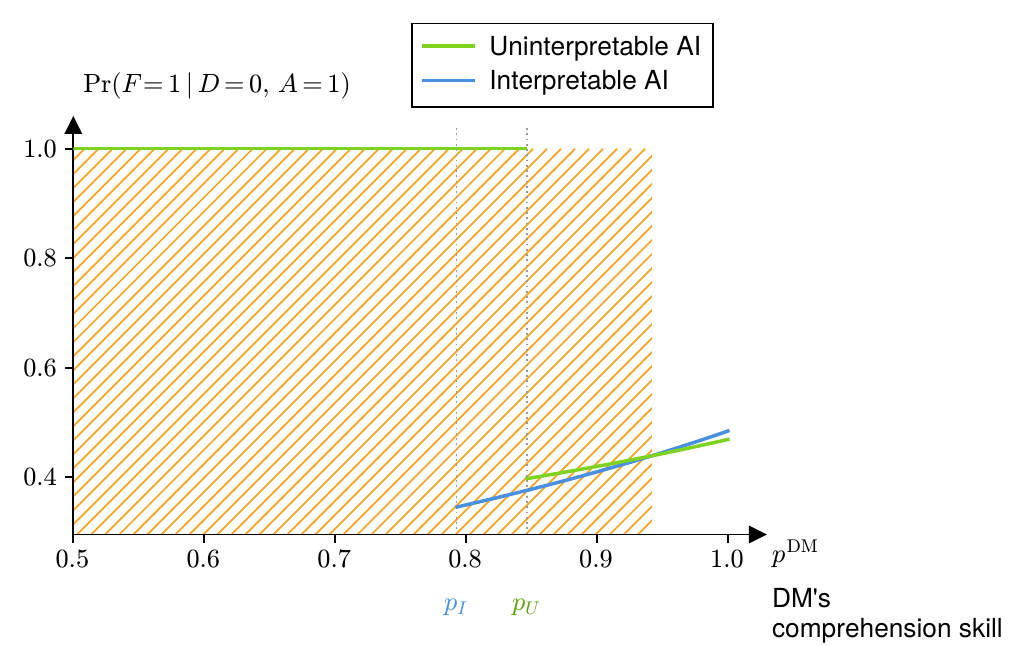}}
    {Conditional Probability of Persuasion After Disagreement. \label{fig: conditional persuasion probability}}
    {Computed at $\gamma=0.247$,
  $\lambda=0.230$, $\pi^\doc=0.819$, $\pi^\ai=0.916$, and $p^\ai=0.775$.}
\end{figure}

So far we have compared the two AIs by which types they persuade. Conditioning on a disagreement, we can instead ask how likely a given type is to be persuaded, namely the probability $\Pr(F=1\mid D=0,A=1)$. Figure~\ref{fig: conditional persuasion probability} plots this probability against $p^\doc$; the shaded region marks the types for which it is weakly higher under an uninterpretable AI than under an interpretable one. Up to $p_U$, the uninterpretable AI persuades with probability one, and strictly more often than the interpretable AI on $(p_I,p_U]$. The shaded region may extend beyond $p_U$, but once $p^\doc$ is large enough the ranking reverses. Intuitively, an interpretable AI still persuades whenever it reveals a positive feature in the DM's critical dimension, whereas an uninterpretable AI offers only a recommendation that a high-comprehension DM may discount. Online Appendix~\ref{Appendix conditional persuasion} makes this comparison precise.

The greater persuasiveness of an uninterpretable AI suggests that lowering
interpretability can facilitate AI adoption. This is broadly consistent with
findings that users sometimes comply more readily when a decision aid provides
fewer explanations. In medicine, \citet{clement2021increasing} observe higher
adherence to anti-rejection drug recommendations when local explanations are
withheld, and \citet{destefano2022providing} report higher compliance with an
uninterpretable inventory algorithm than with a similarly performing interpretable
one.\smallskip

\begin{remark}\label{remark on Proposition 1}
As noted in Remark~\ref{remark on Lemma 2}, Lemma~\ref{lemma I interpretable AI} continues to hold when attention signals can incur both Type~\Rnum{1} and Type~\Rnum{2} errors. In particular, an attention difference remains more persuasive than a comprehension difference. Under this alternative error structure, Proposition~\ref{prop I enhance} therefore follows by the same argument as in this section.

It is also worth noting that if we reverse Assumption~\ref{assumption: signals}, so that attention signals incur no Type~\Rnum{2} errors but may incur Type~\Rnum{1} errors, then analogous calculations imply $p_U \le p_I$. In this case, uninterpretability reduces persuasiveness. Intuitively, under the reversed assumption an attention difference is less persuasive than a comprehension difference, so making the AI uninterpretable lowers persuasiveness by pooling across the two sources of disagreement.
\end{remark}\smallskip

\begin{remark}\label{remark pos to neg}
Beyond this section, Online Appendix~\ref{Appendix: omitted} analyzes the reverse case in
which the DM's initial decision is $D=1$ while the AI recommends $A=0$, using an
analogous classification of attention and comprehension differences. The main findings are as follows. If the AI is
interpretable, persuasion occurs if and only if (\rnum{1}) the DM and the AI have
conflicting comprehension of the critical dimension; (\rnum{2}) the DM does not observe
two positive features; and (\rnum{3}) $p^\doc<\tilde{p}_I$ for some threshold
$\tilde{p}_I$. If the AI is uninterpretable, persuasion requires only condition (\rnum{2})
and $p^\doc<\tilde{p}_U$ for some threshold $\tilde{p}_U$. These thresholds satisfy
$\tilde{p}_U\leq \tilde{p}_I\leq p_I\leq p_U$. This ordering partitions the parameter
space, allowing us to analyze different types of disagreement separately.
\end{remark}\smallskip

Although Proposition~\ref{prop I enhance} shows that a less interpretable AI is more persuasive, it does not follow that it improves decision accuracy. The reason is that an accuracy-maximizing DM weakly benefits from more information, and interpretability gives her more: she sees the AI's signals $(X^\ai,C^\ai)$, not merely its recommendation $A$. Absent any incentive conflict, hiding these signals can only weaken her decision, so the greater persuasiveness of an uninterpretable AI typically comes at an accuracy cost. This cost arises even when the AI and the DM agree.

\begin{proposition}[\textbf{Loss from Uninterpretability under Agreement}]\label{proposition agreement}
Suppose $D=A=0$, $X^\ai(C^\doc)=1$, and $p^\doc>\underline{p}$ for some threshold $\underline{p}$. The DM chooses $F=1$ under an interpretable AI but $F=0$ under an uninterpretable one. Interpretability strictly raises accuracy in this case.
\end{proposition}

We show this formally in \hyperref[Appendix: Proof]{Appendix}. The point of the proposition is that even when the decisions agree, the details behind them may be important for the DM to reverse her decision. The DM and the AI reach $D=A=0$ for different reasons. The DM finds nothing positive in the dimension she regards as critical, while the AI, regarding a different dimension as critical, has nonetheless observed a positive feature in the DM's. Under an interpretable AI, this feature is exposed. Because the DM is sufficiently precise in her comprehension, she then revises her belief toward a positive state and corrects her decision. Under an uninterpretable AI, by contrast, she sees only a recommendation that agrees with her own decision. This conveys nothing she did not already know, and she does not revise her initial decision. The details behind an agreeing recommendation can thus be decisive, and interpretability is what makes them visible.

\section{Career Concerns}\label{sec career}
We have shown that a less interpretable AI is more persuasive but, absent any incentive conflict, reduces decision accuracy. We now turn to settings where an incentive conflict distorts the DM's behavior, so that this cost can instead become a gain: lowering interpretability can make the AI more persuasive and, by curbing the distortion, also improve decision accuracy. We develop this in a leading application, career concerns, in which the DM is rewarded for appearing competent even when doing so sacrifices accuracy. 

Such concerns can slow AI adoption, because openly relying on AI may itself signal lower ability or higher replaceability. \citet{Arkes2007} and \citet{Victoria2013} show that physicians who use computerized decision aids are often viewed as less capable and evaluated more negatively. Similarly, \citet{Reif2025} find that workers who use AI are perceived as more replaceable, and survey evidence links fear of replacement to reluctance to adopt AI \citep{microsoft-wti-2024}. In these settings, reputation and advancement depend on perceived competence, and that perception is shaped by how DMs respond when AI contradicts their initial decisions. Consistent with this, \citet{almog2024ai} document and quantify the costs umpires face when AI overrules them, and \citet{almog2025ai} show that making AI reliance more visible during disagreement reduces AI use. 

In what follows, we study how concerns about reputation for comprehension skill affect a DM's willingness to follow AI recommendations when those recommendations conflict with the DM's initial view.

\subsection{Setup} 
We follow \citet{Li2025} in modeling the DM's career concerns. The DM can be one of two types, distinguished by their comprehension skills. A high-type DM has perfect comprehension ($p^{\doc_H} = 1$), while a low-type DM has imperfect comprehension ($p^{\doc_L} < 1$). Both types share the same attention skill, $\pi^\doc\in(0,1)$. Ex ante, the DM is high-type with probability $\tau$ and low-type with probability $1 - \tau$. Each DM privately knows her own type. We also assume:

\begin{assumption}\label{assumption career concerns}
$\tilde{p}_I < p^{\doc_L} < p^\ai$ and $p_U=1$, where $p_U$ is the threshold for the DM to be persuaded by an uninterpretable AI to choose a positive final decision (see Proposition~\ref{prop I enhance}), and $\tilde{p}_I$ is the threshold for the DM to be persuaded by an interpretable AI to choose a negative final decision (see Remark~\ref{remark pos to neg} and Online Appendix~\ref{Appendix: omitted}).
\end{assumption}

This assumption has three implications. First, when the DM initially chooses $D=0$ and the AI recommends $A=1$, an uninterpretable AI persuades both types, including the high type.\footnote{We impose $p_U=1$ only to ensure that $p^{\doc_H}\leq p_U$. If instead $p^{\doc_H}<1$, we could allow $p_U<1$ as well.} Second, when the DM initially chooses $D=1$ and the AI recommends $A=0$, the AI persuades neither type. Third, the low-type DM has weaker comprehension than the AI. Together, these conditions capture a setting where career concerns are most salient: the AI can improve the low type's decisions through better comprehension, but deferring to it may be resisted because it signals low ability. The \hyperref[Appendix: Proof]{Appendix} verifies that these conditions can hold at the same time.

The DM maximizes reputation rather than decision accuracy. We define reputation as an evaluator's posterior belief that the DM is high-type. The evaluator always observes the DM's signals $X^\doc$ and $C^\doc$, as well as her initial decision $D$ and final decision $F$. If the AI is interpretable, the evaluator also observes the AI's signals $X^\ai$ and $C^\ai$ and its recommendation $A$. If the AI is uninterpretable, the evaluator observes only $A$ from the AI.

We abstract from any self-reporting problem by assuming the DM's signals are directly observed. This captures settings in which the DM must document her analysis before seeing the AI's recommendation, so truthful reporting is optimal. Because agreement with the AI improves reputation once the AI's signals are revealed, a career-concerned DM has no incentive to misreport.

We use Perfect Bayesian Equilibrium as our solution concept. An \emph{efficient}
decision maximizes decision accuracy given the DM's information at the time of the
decision. Following \citet{Li2025}, we focus on equilibria in which the high-type
DM makes efficient decisions, and we study how career concerns distort the
low-type DM's behavior and generate inefficiencies. We adopt the off-path belief
that any final decision departing from the high-type DM's efficient action is
attributed to the low-type DM. This belief is obtained by perturbing the DM's
payoff to place weight $\delta>0$ on decision accuracy and letting $\delta\to 0$,
so that it arises as a limit of beliefs derived from Bayes' rule.

\subsection{Analysis and Results} 
To illustrate how career concerns can lead to inefficient decisions, consider disagreement driven by a comprehension difference. Suppose the DM and the AI observe the same features, $X^\doc=X^\ai=(0,1)$, but disagree about which dimension is critical: the DM believes it is the left one ($C^\doc=L$), whereas the AI believes it is the right one ($C^\ai=R$). In this situation, the AI's comprehension yields a positive recommendation ($A=1$), while the high-type DM's comprehension yields a negative initial decision, as in Lemma~\ref{lem benchmark}. To protect her reputation, the low-type DM mimics the high type by also choosing a negative decision. As a result, both types initially choose $D=0$.

Without career concerns, both types would resolve this disagreement efficiently, meaning that each type maximizes decision accuracy given what she observes. By Lemma~\ref{lemma I interpretable AI}, decision accuracy is maximized when the high-type DM keeps her negative decision ($F=0$), since she has perfect comprehension, and when the low-type DM defers to the AI's better comprehension and switches to a positive decision ($F=1$). With career concerns, however, the low-type DM may instead deviate from the efficient choice. How large the distortion is depends on the AI's interpretability.

When the AI is interpretable and the evaluator observes all signals and decisions, the DM's response to disagreement can reveal her type. If each type played the efficient strategy, with the high type sticking to $F=0$ and the low type switching to $F=1$, Bayes' rule would imply
\begin{align}
\Pr(\text{DM}_H\mid F=0,\comp,\text{interpretable AI},\text{efficient}) &= 1,\\
\Pr(\text{DM}_H\mid F=1,\comp,\text{interpretable AI},\text{efficient}) &= 0.
\end{align}
Anticipating this reputational gap, the low type deviates from the efficient strategy and mimics the high type by also choosing $F=0$. In equilibrium,
\begin{equation}
\Pr(\text{DM}_H\mid F=0,\comp,\text{interpretable AI})=\tau > \Pr(\text{DM}_H\mid F=1,\comp,\text{interpretable AI})=0.
\end{equation}
Thus, interpretability creates a reputation penalty for following the AI in this disagreement, leading the low type to ignore valuable recommendations.

By contrast, when the AI is uninterpretable, the evaluator cannot observe the source of disagreement. Because uninterpretability increases the AI's persuasiveness, even the high-type DM chooses the positive decision, and the low-type DM again mimics the high type. In equilibrium,
\begin{equation}
\Pr(\text{DM}_H\mid F=1,\text{uninterpretable AI})=\tau>\Pr(\text{DM}_H\mid F=0,\text{uninterpretable AI})=0.
\end{equation}
Thus, uninterpretability removes the reputation penalty for following the AI. In particular, the low-type DM can follow the AI's recommendation and choose $F=1$ without revealing her type. She then benefits from the AI's superior comprehension rather than relying on her own.

Behind this contrast is how the evaluator infers the DM's ability, and in equilibrium she is never fooled. Under an interpretable AI the comprehension difference is visible, so the efficient strategy would make resisting the AI signal strong comprehension and following it weak; to avoid signaling weak comprehension, the low type mimics the high type, both pool on resisting, and the posterior stays at the prior $\tau$. Under an uninterpretable AI the source is hidden, both types pool on following, and the posterior is again $\tau$. Uninterpretability therefore does not deceive the evaluator but changes the action on which the types pool. By coarsening the public information, it removes the evidence that would have supported sharper inferences in both directions: not only the reputation penalty that following the AI would otherwise carry, but also the favorable inference that resisting it would otherwise generate.

The next lemma generalizes this example. We analyze how AI interpretability affects whether the low-type DM follows the AI's recommendation. 

\begin{lemma}\label{lemma follow AI or not}
Suppose that disagreement is given by $D=0$ and $A=1$. When the AI is interpretable, the low-type DM follows the AI if the disagreement stems from attention differences but not if it stems from comprehension differences. When the AI is uninterpretable, the low-type DM always follows the AI.
\end{lemma}

Lemma~\ref{lemma follow AI or not} shows that with an interpretable AI, the low-type DM's willingness to follow the recommendation depends on the source of disagreement. With an uninterpretable AI, the source of disagreement is no longer relevant, and the low-type DM always follows the AI.

The logic follows the earlier example. Any final decision that departs from the high-type DM's efficient choice reveals the DM to be low type. The low type therefore follows the AI only when doing so is also efficient for the high type. With an interpretable AI, this is true only when disagreement stems from an attention difference. With an uninterpretable AI, it is true regardless of the source of disagreement. Hence, the low type follows an interpretable AI only under attention differences, and always follows an uninterpretable AI.

The earlier example also shows how making the AI uninterpretable can help the DM make better use of its recommendations. The next proposition studies the welfare implications of this interpretability design.

\begin{proposition}[\textbf{Uninterpretability Improves Accuracy}]\label{prop reputation concerns}
    There exists $\overline{\tau}\in(0,1)$ such that if $\tau<\overline{\tau}$, the accuracy of the final decision is higher when the AI is uninterpretable than when it is interpretable.
\end{proposition}

Proposition~\ref{prop reputation concerns} shows that AI uninterpretability can improve decision accuracy when DMs have career concerns. With an interpretable AI, these concerns can lead a low-type DM to reject the AI recommendation under comprehension differences, even though doing so always reduces her decision accuracy. With an uninterpretable AI, the reputational incentive no longer induces this behavior. Because of the averaging effect, the uninterpretable AI can persuade even the high-type DM, so deferring to the AI does not reveal the DM's type. The low-type DM can then rely on the AI's better comprehension without signaling low ability, which raises her accuracy relative to the interpretable case.\footnote{Uninterpretability also prevents the DM from using the AI's attention signal. When $p^{\doc_L}<p^\ai$, the gain from relying on the AI's better comprehension outweighs this cost (see Lemma~\ref{lem reputation concerns} and the proof of Proposition~\ref{prop reputation concerns} in the \hyperref[Appendix: Proof]{Appendix}).} If the ex ante probability of being a low type is high enough ($\tau<\overline{\tau}$), average decision accuracy across DMs also increases. This result provides a potential explanation for the finding that less interpretable AI can deliver both higher acceptance and higher task performance than an interpretable counterpart with similar accuracy \citep{destefano2022providing}. 

Proposition~\ref{prop reputation concerns} suggests that human–AI collaboration can improve when the system is less transparent about why it disagrees with the decision maker (DM). In radiology, for example, the system can report an overall risk score rather than flagging visual features the physician has already inspected and dismissed as noise. Similarly, in various decision scenarios, what is often needed is a composite prediction or ranking, not a detailed rationale that directly contradicts the DM’s judgment.

This design choice can be valuable in settings where AI use must be disclosed or recorded.\footnote{For instance, Texas Senate Bill 1188 requires a health care practitioner who uses AI for diagnostic purposes to disclose that use to the practitioner’s patients. New York City Administrative Code 20-871 requires employment agencies that use an automated employment decision tool to screen candidates to give advance notice that the tool will be used and to disclose the job qualifications and characteristics the tool will assess.} When the DM cannot hide the fact that they consulted AI, overly transparent explanations can turn a change of mind into an admission of poor judgment. By keeping the recommendation visible while making the underlying reasons opaque, the system allows the DM to incorporate the AI’s information without suffering the reputation loss. 

It is worth noting that the accuracy gain from uninterpretability depends on the low type being able to mimic the high type's final decision. When the high-type action is instead difficult to imitate, for instance because decisions are verified ex post or subject to audit and documentation requirements, resisting a comprehension difference becomes a favorable signal that the low type cannot replicate. The high type can then separate in equilibrium, and in this case making the AI uninterpretable may no longer improve accuracy.

\subsection{Transparency Policies}\label{sec transparency policies}
Beyond interpretability design, two transparency policies may help manage DMs' career concerns and affect whether uninterpretability improves decision accuracy. The first policy is to reveal the correct decision after the DM's final decision, with the aim of encouraging better use of available information. In our baseline setting, this policy has no effect because decision accuracy does not enter the DM's payoff. In Online Appendix~\ref{appendix Transparency on Correct Decisions}, we study a more general model in which the DM's payoff is a weighted average of decision accuracy and reputation. We show that (i) when the AI is interpretable, revealing ex post correctness can weakly improve decision accuracy, and (ii) when career concerns are sufficiently strong, making the AI uninterpretable can still be beneficial (Proposition~\ref{prop transparency on correct decisions}).

The intuition is as follows. When the correct decision is revealed, the evaluator can partly base the DM's reputation on whether her final decision was correct. This reduces the low-type DM's incentive to mimic the high type, as such mimicry is inefficient and results in lower accuracy. For example, when an interpretable AI contradicts her comprehension, the low-type DM becomes more inclined to follow the AI rather than persist in a likely mistake. This improves decision accuracy.

However, as career concerns intensify, this effect weakens. When reputation motives dominate the DM's objectives, the low-type DM will always mimic the high type, so revealing the correct decision has no effect. In this case, making AI uninterpretable can still improve decision accuracy by making it reputationally safe for a career-concerned DM to follow the AI’s recommendation.

The second policy to mitigate DMs' career concerns is to allow private use of AI, ensuring that neither the AI's recommendation nor the DM's signals and initial decision are observable to the evaluator. Under this policy, the DM reviews all available information privately and then makes a single public decision. In Online Appendix~\ref{appendix Transparency on the Influence of AI}, Proposition~\ref{prop transparency on the influence of AI} demonstrates that this policy can improve decision accuracy. By keeping the DM's response to the AI private, she can better utilize the AI's information and make more efficient decisions.

\subsection{Extensions}

\noindent\textbf{Behavioral elements.} Our model does not incorporate behavioral elements; thus far, we have assumed that information processing follows standard Bayesian updating. However, this does not mean that our model is incompatible with bounded rationality on the part of the DM. For example, \citet{Agarwal2023} document that radiologists exhibit overconfidence by underweighting AI predictions. Similarly, \citet{Caplin25abc} find that human decision-makers realize only modest gains from AI assistance because they overestimate their own accuracy. Such overconfidence can be incorporated into our model by assuming that the DM is overconfident in her comprehension. In this case, the gap in persuasiveness between attention differences and comprehension differences becomes even larger, further strengthening the enhancing effect of AI's uninterpretability.\medskip

\noindent\textbf{Effort incentives.} In addition to the career concerns studied earlier, other incentive conflicts may also lead an uninterpretable AI to outperform an interpretable one in terms of decision accuracy. One such conflict concerns effort provision. Suppose that the DM can incur a cost to obtain an additional attention signal before making her final decision. Will she exert effort, and how does interpretability influence this choice? Online Appendix~\ref{Appendix moral hazard} addresses this question. We show that making the AI uninterpretable can increase the DM's incentive to exert effort by raising the perceived value of the additional signal. This, in turn, improves the accuracy of her decision.\medskip 

\section{Conclusion}
This paper develops a framework for how AI persuades human DMs when the two disagree. We emphasize two distinct sources of disagreement. The first is an attention difference, where the AI surfaces relevant features the DM overlooked. The second is a comprehension difference, where both observe the same features but disagree about how to interpret or weight them. The former typically supports collaboration, whereas the latter often triggers resistance due to career concerns.

Our central finding is that the design of the AI's information structure, specifically its interpretability, fundamentally alters how DMs attribute the source of disagreement. We show that uninterpretable AI can be more persuasive by pooling these sources of disagreement. By preventing the DM from knowing that a disagreement is purely a matter of judgment, uninterpretability leads the DM to attribute the conflict to potential attention oversights, which are more persuasive. This ``averaging effect'' is particularly powerful when DMs have low attention skills. When DMs are career concerned, making AI uninterpretable can even improve welfare.

Our results have significant implications for the management of AI integration in organizations. As AI tools become more prevalent in knowledge work, managers must recognize that barriers to adoption are not solely cognitive but also social and reputational. The fear of replacement or devaluation is a powerful deterrent. Therefore, the design of AI systems should not be viewed purely as a technical challenge of maximizing accuracy or transparency, but as an organizational design challenge of aligning incentives.

Specifically, our analysis suggests that organizations might benefit from strategic opacity. Rather than defaulting to full explainability, firms could design protocols where AI provides ``flags'' (resembling attention signals) rather than detailed competing rationales (comprehension signals) that challenge a professional's status. Furthermore, creating environments where ``private consultation'' with AI is possible can mitigate the image concerns that drive resistance, allowing DMs to incorporate AI information more effectively.

\phantomsection
\label{Appendix: Proof}
\pdfbookmark[1]{Appendix. Proofs}{appendix-proofs}
\begin{APPENDIX}{Proofs}

\subsection*{Proof of Lemma~\ref{lem benchmark}}
Without loss of generality, suppose $C^\doc=L$. We compute the DM's posterior by marginalizing over the critical dimension $C$ and the noise $\varepsilon$:
\begin{equation}\label{eq: marginalization}
\Pr(X^\doc,C^\doc,Z=z)=\Pr(Z=z)\sum_{C\in\{L,R\}}\sum_{\varepsilon\in\{0,1\}}\Pr(C)\,\Pr(\varepsilon)\,\Pr(X^\doc,C^\doc\mid Z=z,C,\varepsilon),
\end{equation}
and the likelihood ratio of the state is the quotient of \eqref{eq: marginalization} at $z=1$ and $z=0$. Given $(Z,C,\varepsilon)$, the features are determined by \eqref{eq DGP}, and the DM's signal $X^\doc$ is then conditionally independent across dimensions, with no Type~\Rnum{1} error and Type~\Rnum{2} rate $1-\pi^\doc$. There are four cases for $X^\doc$.

In case (\rnum{1}), $X^\doc=(1,1)$. Since attention signals admit no Type~\Rnum{1} error, both features are positive, so $Z=1$ with certainty and $\Pr(Z=1\mid X^\doc,C^\doc)=1$. Thus $D=1$.

In case (\rnum{2}), $X^\doc=(0,0)$. The DM observes no positive feature, so every $(C,\varepsilon)$ is consistent with the data under both states; the two states differ only in whether the critical feature is a positive that the DM missed. The common configuration cancels from \eqref{eq: marginalization}, leaving
\[
\frac{\Pr(Z=1\mid X^\doc,C^\doc)}{\Pr(Z=0\mid X^\doc,C^\doc)}=\frac{\gamma(1-\pi^\doc)}{1-\gamma}<1,
\]
where the inequality follows from $\gamma<\tfrac12$. Thus $D=0$.

In case (\rnum{3}), $X^\doc=(1,0)$. Here $X^\doc(L)=1$ forces $X(L)=1$, while $X^\doc(R)=0$ leaves $X(R)\in\{0,1\}$, so the sum in \eqref{eq: marginalization} runs over whether $R$ carries a missed positive. This gives
\[
\frac{\Pr(Z=1\mid X^\doc,C^\doc)}{\Pr(Z=0\mid X^\doc,C^\doc)}=\frac{\gamma[p^\doc(1-\lambda\pi^\doc)+(1-p^\doc)\lambda(1-\pi^\doc)]}{(1-\gamma)(1-p^\doc)\lambda}\geq 1,
\]
where the inequality follows from $\gamma(1-\lambda\pi^\doc)\geq(1-\gamma)\lambda$ and $p^\doc\geq 1-p^\doc$. Thus $D=1$.

In case (\rnum{4}), $X^\doc=(0,1)$. Now $X^\doc(R)=1$ forces $X(R)=1$, while $X^\doc(L)=0$ leaves $X(L)\in\{0,1\}$, so the sum runs over whether $L$ carries a missed positive. This gives
\[
\frac{\Pr(Z=1\mid X^\doc,C^\doc)}{\Pr(Z=0\mid X^\doc,C^\doc)}=\frac{\gamma[p^\doc(1-\pi^\doc)\lambda+(1-p^\doc)(1-\lambda\pi^\doc)]}{(1-\gamma)p^\doc\lambda}<1,
\]
where the inequality follows from Assumption~\ref{assumption gamma lambda 2}. Thus $D=0$.\qed\medskip

\subsection*{Proof of Lemma~\ref{lemma I interpretable AI}}
According to Lemma~\ref{lem benchmark}, $X^\doc(C^\doc)=0$ and $X^\ai(C^\ai)=1$ when $D=0$ and $A=1$. Since the DM's and the AI's signals are independent conditional on the fundamentals $(Z,C,\varepsilon)$, the joint distribution marginalizes as
\begin{equation}\label{eq: marginalization two}
\begin{aligned}
\Pr(X^\doc,C^\doc,X^\ai,C^\ai,Z=z)
={}&\Pr(Z=z)\sum_{C\in\{L,R\}}\sum_{\varepsilon\in\{0,1\}}\Pr(C)\Pr(\varepsilon)\\
&\times\Pr(X^\doc,C^\doc\mid Z=z,C,\varepsilon)\,\Pr(X^\ai,C^\ai\mid Z=z,C,\varepsilon),
\end{aligned}
\end{equation}
and the likelihood ratio of the state is the quotient of \eqref{eq: marginalization two} at $z=1$ and $z=0$. As in \eqref{eq: marginalization}, each agent's factor is determined by \eqref{eq DGP} and carries no Type~\Rnum{1} error.
Suppose first that $X^\ai(C^\doc)=1$. This event arises both when $C^\ai=C^\doc$ and when $C^\ai\neq C^\doc$ (for which $X^\ai=(1,1)$); both are covered below, since each reveals a positive feature in the DM's critical dimension. By \eqref{eq: marginalization two}, the likelihood ratio of the state is
\begin{align}\label{eq atten LR}
    &\frac{\Pr(Z=1\mid  X^\doc, C^\doc, X^{\ai}, C^\ai )}{\Pr(Z=0\mid  X^\doc, C^\doc, X^{\ai}, C^\ai )}\\
    &\qquad=\begin{dcases}
        +\infty, \text{ if } \left(\max\{X^\doc(L),X^{\ai}(L)\},\max\{X^\doc(R),X^{\ai}(R)\}\right)=(1,1),\\
        \frac{\gamma[p^\doc p^\ai(1-\lambda)+\lambda(1-\pi^\doc)(1-\pi^\ai)(1-p^\doc-p^\ai+2p^\doc p^\ai)]}{(1-\gamma)(1-p^\doc)(1-p^\ai)\lambda}, \text{ otherwise. }
    \end{dcases}
\end{align}
In the first case, both observed positives are conclusive since attention signals carry no Type~\Rnum{1} error, so $Z=1$ and the ratio is $+\infty$. In the second, $\gamma(1-\lambda)\geq(1-\gamma)\lambda$ and $p^\doc p^\ai\geq(1-p^\doc)(1-p^\ai)$ make the ratio weakly greater than one. Either way, $F=1$.
Now suppose $X^\ai(C^\doc)=0$. Then $C^\ai\neq C^\doc$, and the AI's positive recommendation reflects a positive feature in its own critical dimension. By \eqref{eq: marginalization two}, the likelihood ratio of the state is
\begin{equation}\label{eq comp LR}
    \frac{\gamma[p^\ai(1-p^\doc)(1-\lambda)+\lambda(1-\pi^\doc)(1-\pi^\ai)(p^\doc+p^\ai-2p^\doc p^\ai)]}{(1-\gamma)p^\doc(1-p^\ai)\lambda},
\end{equation}
which is at least one if and only if 
\begin{equation}\label{eq comp threshold}
    p^\doc\leq p^\ai\cdot\frac{(1-\lambda)/\lambda+(1-\pi^\doc)(1-\pi^\ai)}{(1-p^\ai)(1-\gamma)/\gamma+p^\ai(1-\lambda)/\lambda+(2p^\ai-1)(1-\pi^\doc)(1-\pi^\ai)}.
\end{equation}
Denote the right-hand side as $p_I$. We conclude that when $X^\ai(C^\doc)=0$, $F=1$ if and only if $p^\doc\leq p_I$. To verify that $p_I$ is increasing in $p^\ai$, write
\[
p_I=\frac{p^\ai(a+b)}{(1-p^\ai)g+p^\ai a+(2p^\ai-1)b},\qquad\text{where } a=\tfrac{1-\lambda}{\lambda},\ b=(1-\pi^\doc)(1-\pi^\ai),\ g=\tfrac{1-\gamma}{\gamma}.
\]
Then
\[
\frac{\partial p_I}{\partial p^\ai}=\frac{(a+b)(g-b)}{[(1-p^\ai)g+p^\ai a+(2p^\ai-1)b]^2}>0,
\]
since $g-b=\tfrac{1-\gamma}{\gamma}-(1-\pi^\doc)(1-\pi^\ai)>0$ by Assumption~\ref{assumption gamma lambda 1}.\qed\medskip

\subsection*{Proof of Corollary~\ref{corollary 1}}
Suppose that an attention difference occurs. Decompose the conditional probability that the state is positive as:
\begin{align}
\Pr(Z=1\mid \atten,\cI^\doc)=&\Pr(Z=1\mid \cE,\atten,\cI^\doc)\cdot\Pr(\cE\mid \atten,\cI^\doc)\\
&+\Pr(Z=1\mid \cF,\atten,\cI^\doc)\cdot\Pr(\cF\mid \atten,\cI^\doc),
\end{align}
where $\cE$ denotes the event that $(\max\{X^\doc(L),X^{\ai}(L)\},\max\{X^\doc(R),X^{\ai}(R)\})=(1,1)$, and $\cF$ denotes the event that complements $\cE$. According to \eqref{eq atten LR}, $\Pr(Z=1\mid \cE,\atten,\cI^\doc)=1$ and $\Pr(Z=1\mid \cF,\atten,\cI^\doc)\geq\frac{1}{2}$. Since $\Pr(\cE\mid \atten,\cI^\doc)$ and $\Pr(\cF\mid \atten,\cI^\doc)$ are strictly positive, $\Pr(Z=1\mid \atten,\cI^\doc)>\frac{1}{2}$. 

Then, suppose that a comprehension difference occurs. According to \eqref{eq comp LR} and \eqref{eq comp threshold}, $\Pr(Z=1\mid \comp,\cI^\doc)\geq\frac{1}{2}$ if and only if $p^\doc\leq p_I$.\qed\medskip

\subsection*{Proof of Proposition~\ref{prop I enhance}}
Without loss of generality, suppose $C^\doc=L$. By Lemma~\ref{lem benchmark}, $D=0$
implies $X^\doc(L)=0$, so the DM's attention signal is either (\rnum{1}) $X^\doc=(0,0)$
or (\rnum{2}) $X^\doc=(0,1)$.

Since the AI is uninterpretable, the DM observes $A$ but not $(X^\ai,C^\ai)$. Relative
to \eqref{eq: marginalization two}, we marginalize over the AI's hidden signals through
its recommendation:
\begin{equation}\label{eq: marginalization uninterpretable}
\begin{aligned}
\Pr(X^\doc,C^\doc,A=1,Z=z)
={}&\Pr(Z=z)\sum_{C\in\{L,R\}}\sum_{\varepsilon\in\{0,1\}}\Pr(C)\Pr(\varepsilon)\\
&\times\Pr(X^\doc,C^\doc\mid Z=z,C,\varepsilon)\,\Pr(A=1\mid Z=z,C,\varepsilon).
\end{aligned}
\end{equation}
Given $(Z=z,C,\varepsilon)$, the AI recommends $A=1$ if and only if it observes a positive
feature in its own critical dimension, so
\begin{equation}\label{eq: A positive marginal}
\begin{aligned}
\Pr(A=1\mid Z=z,C,\varepsilon)
&=\sum_{c\in\{C,-C\}}\Pr(C^\ai=c\mid C)\,\Pr(X^\ai(c)=1\mid Z=z,C,\varepsilon)\\
&=\pi^\ai\bigl[p^\ai z+(1-p^\ai)\varepsilon\bigr].
\end{aligned}
\end{equation}
The likelihood ratio of the state is the quotient of
\eqref{eq: marginalization uninterpretable} at $z=1$ and $z=0$; the common factor
$\pi^\ai$ cancels.

By \eqref{eq: A positive marginal}, if $p^\ai=1$ then $\Pr(A=1\mid Z=0,C,\varepsilon)=0$,
so $A=1$ reveals a positive feature in the true critical dimension; hence $Z=1$ with
certainty, the DM sets $F=1$ for every $p^\doc$, and the claim holds with $p_U=1$. We
therefore suppose $p^\ai<1$ below, so the denominators that follow are positive.

\emph{Case (\rnum{1}): $X^\doc=(0,0)$.} By
\eqref{eq: marginalization uninterpretable}--\eqref{eq: A positive marginal},
\[
\frac{\Pr(Z=1\mid \cI^\doc)}{\Pr(Z=0\mid \cI^\doc)}
=\frac{\gamma\bigl[(1-\lambda)p^\ai+\lambda(1-\pi^\doc)\bigr]}{(1-\gamma)(1-p^\ai)\lambda}
\geq 1,
\]
where the inequality uses $\gamma(1-\lambda)\geq(1-\gamma)\lambda$ (equivalently
$\gamma\geq\lambda$, Assumption~\ref{assumption gamma lambda 1}) and $p^\ai\geq\tfrac12$.
This ratio does not depend on $p^\doc$, so $F=1$ for every $p^\doc$.

\emph{Case (\rnum{2}): $X^\doc=(0,1)$.} By
\eqref{eq: marginalization uninterpretable}--\eqref{eq: A positive marginal},
\begin{equation}\label{eq likelihood ratio proof prop 1}
\frac{\Pr(Z=1\mid \cI^\doc)}{\Pr(Z=0\mid \cI^\doc)}
=\frac{\gamma\bigl[\lambda(1-\pi^\doc)+(1-\lambda)(1-p^\doc)p^\ai\bigr]}
{(1-\gamma)p^\doc(1-p^\ai)\lambda}.
\end{equation}
The numerator is decreasing in $p^\doc$ (its derivative is $-(1-\lambda)p^\ai\leq0$) and
the denominator is increasing, so the ratio is decreasing in $p^\doc$. Hence it is at least
one if and only if $p^\doc\leq\hat p_U$, where
\[
\hat p_U:=\frac{(1-\lambda)p^\ai/\lambda+(1-\pi^\doc)}
{(1-\gamma)(1-p^\ai)/\gamma+(1-\lambda)p^\ai/\lambda}.
\]
Because $p^\doc\leq1$, the persuasion threshold is $p_U:=\min\{1,\hat p_U\}\leq1$, and in
this case $F=1$ if and only if $p^\doc\leq p_U$. (If $\hat p_U\geq1$ then $p_U=1$ and the
uninterpretable AI persuades every admissible type.)

Combining Cases (\rnum{1}) and (\rnum{2}), the uninterpretable AI persuades the DM whenever
$p^\doc\leq p_U$, proving Part~\ref{part:uninterp-thres}.

\emph{Averaging effect (Part~\ref{part:uninterp-avg}).} Fix $X^\doc=(0,1)$ and set
$p^\doc=p_I$. By Corollary~\ref{corollary 1},
\[
\Pr(Z=1\mid \atten,\cI^\doc)>\Pr(Z=1\mid \comp,\cI^\doc)=\tfrac12,
\]
so \eqref{eq: expression} gives $\Pr(Z=1\mid \cI^\doc)\geq\tfrac12$; equivalently the ratio
in \eqref{eq likelihood ratio proof prop 1} is at least one at $p^\doc=p_I$. Since that
ratio equals one at $p^\doc=\hat p_U$ and is decreasing in $p^\doc$, it follows that
$\hat p_U\geq p_I$, with $\hat p_U>p_I$ whenever $\pi^\doc<1$ (then
$\Pr(\atten\mid \cI^\doc)>0$). Moreover $p_I\leq1$: by \eqref{eq comp threshold} and
$p^\ai<1$, this reduces to $(1-\pi^\doc)(1-\pi^\ai)\leq(1-\gamma)/\gamma$, which holds
since $(1-\pi^\doc)(1-\pi^\ai)\leq1<(1-\gamma)/\gamma$ (as $\gamma<\tfrac12$). Hence
$p_U=\min\{1,\hat p_U\}\geq p_I$. At $p^\ai=1$ one has $p_I=1=p_U$, so the inequality is
strict precisely when $\pi^\doc<1$ and $p^\ai<1$.

\emph{Attribution effect (Part~\ref{part:uninterp-att}).} We show $\hat p_U-p_I$ is
decreasing in $\pi^\doc$; on the range where $p_U<1$ (equivalently $\hat p_U<1$, so
$p_U=\hat p_U$) the gap $p_U-p_I$ then increases as $\pi^\doc$ decreases.

For intuition, a less attentive DM places more weight on the attention difference. With
$w:=1-\pi^\doc$ and $S:=p^\doc+p^\ai-2p^\doc p^\ai\geq0$,
\[
\frac{\Pr(\atten\mid \cI^\doc)}{\Pr(\comp\mid \cI^\doc)}=\frac{\alpha\,w}{\beta+\delta\,w},
\]
where
\[
\begin{aligned}
\alpha&=\gamma\lambda\bigl[p^\doc p^\ai+(1-p^\doc)(1-p^\ai)+\pi^\ai S\bigr],\\
\beta&=\gamma(1-\lambda)p^\ai(1-p^\doc)+\lambda(1-\gamma)p^\doc(1-p^\ai),\\
\delta&=\gamma\lambda(1-\pi^\ai)S,
\end{aligned}
\]
with $\alpha,\beta>0$ and $\delta\geq0$. As $\tfrac{\alpha w}{\beta+\delta w}$ is increasing
in $w$, the odds decrease in $\pi^\doc$, so $\Pr(\atten\mid \cI^\doc)$ decreases in
$\pi^\doc$.

For the comparative static, set
\[
\begin{aligned}
N_I&:=\Bigl(\tfrac{1-\gamma}{\gamma}+\tfrac{1-\lambda}{\lambda}\Bigr)
      p^\ai(1-p^\ai)(1-\pi^\ai),\\
K_U&:=(1-p^\ai)\tfrac{1-\gamma}{\gamma}+p^\ai\tfrac{1-\lambda}{\lambda},\\
K_I&:=(2p^\ai-1)(1-\pi^\doc)(1-\pi^\ai),
\end{aligned}
\]
so $K_U+K_I$ is the denominator of $p_I$ in \eqref{eq comp threshold}. Differentiating
$p_I$ and $\hat p_U$,
\[
\frac{\partial p_I}{\partial\pi^\doc}=-\frac{N_I}{(K_U+K_I)^2},\qquad
\frac{\partial \hat p_U}{\partial\pi^\doc}=-\frac{1}{K_U}.
\]
Now $K_I\geq0$ and
\[
N_I\leq\Bigl(\tfrac{1-\gamma}{\gamma}+\tfrac{1-\lambda}{\lambda}\Bigr)p^\ai(1-p^\ai)
\leq K_U\leq K_U+K_I,
\]
the middle inequality because
\[
K_U-\Bigl(\tfrac{1-\gamma}{\gamma}+\tfrac{1-\lambda}{\lambda}\Bigr)p^\ai(1-p^\ai)
=(1-p^\ai)^2\tfrac{1-\gamma}{\gamma}+(p^\ai)^2\tfrac{1-\lambda}{\lambda}\geq0.
\]
Hence
\[
\frac{\partial \hat p_U}{\partial\pi^\doc}=-\frac{1}{K_U}
\leq-\frac{N_I}{(K_U+K_I)^2}=\frac{\partial p_I}{\partial\pi^\doc}\leq0,
\]
so $\hat p_U-p_I$ is decreasing in $\pi^\doc$. On the range where $p_U<1$ we have
$p_U=\hat p_U$, so $p_U-p_I$ increases as $\pi^\doc$ decreases. Once $\hat p_U\geq1$ the cap
binds, $p_U=1$, and the uninterpretable AI already persuades every admissible
type.\qed\medskip

\subsection*{Proof of Proposition~\ref{proposition agreement}}
Without loss of generality, suppose $C^\doc=L$. On the event $D=A=0$ and $X^\ai(C^\doc)=1$, Lemma~\ref{lem benchmark} implies $X^\doc(L)=0$ and $X^\ai(C^\ai)=0$. Since $X^\ai(L)=1$, this further implies $C^\ai=R$ and $X^\ai(R)=0$. The only remaining free signal is the DM's observation in her non-critical dimension, $X^\doc(R)\in\{0,1\}$, and we treat the two values in turn.

\emph{Interpretable AI.} First suppose $X^\doc(R)=1$. Then $X^\ai(L)=X^\doc(R)=1$, and since attention signals have no Type~\Rnum{1} error by Assumption~\ref{assumption: signals}, both underlying features are positive. Hence $Z=1$ with certainty and $F=1$, regardless of $p^\doc$.

Next suppose $X^\doc(R)=0$, so the DM observes $X^\doc=(0,0)$. Writing $\kappa:=(1-\pi^\doc)(1-\pi^\ai)$ and marginalizing over $C$ and $\varepsilon$,
\begin{align}
&\Pr\big(Z=1,X^\doc=(0,0),C^\doc=L,X^\ai=(1,0),C^\ai=R\big)\\
&\qquad=\tfrac{\gamma}{2}\pi^\ai(1-\pi^\doc)\big[p^\doc(1-p^\ai)(1-\lambda+\lambda \kappa)+(1-p^\doc)p^\ai\lambda \kappa\big],\\
&\Pr\big(Z=0,X^\doc=(0,0),C^\doc=L,X^\ai=(1,0),C^\ai=R\big)\\
&\qquad=\tfrac{1-\gamma}{2}(1-p^\doc)p^\ai\lambda\pi^\ai(1-\pi^\doc).
\end{align}
The likelihood ratio of the state is therefore
\begin{align}
\frac{\gamma\big[p^\doc(1-p^\ai)(1-\lambda+\lambda \kappa)+(1-p^\doc)p^\ai\lambda \kappa\big]}{(1-\gamma)(1-p^\doc)p^\ai\lambda},
\end{align}
which is weakly greater than one if and only if
\begin{align}\label{eq: p underline}
p^\doc\geq\underline{p}:=\frac{p^\ai\lambda\big(\tfrac{1-\gamma}{\gamma}-\kappa\big)}{(1-p^\ai)(1-\lambda+\lambda \kappa)+p^\ai\lambda\big(\tfrac{1-\gamma}{\gamma}-\kappa\big)}.
\end{align}
Thus, under an interpretable AI, the DM chooses $F=1$ whenever $p^\doc\geq\underline{p}$ (and for any $p^\doc$ when $X^\doc(R)=1$). The threshold lies in the admissible range, and is interior in the nondegenerate case. Since $\kappa\leq 1$ and $\gamma<\tfrac12$ (Assumption~\ref{assumption gamma lambda 1}), $\tfrac{1-\gamma}{\gamma}-\kappa\geq\tfrac{1-2\gamma}{\gamma}>0$, so the numerator of $\underline{p}$ is positive. The denominator exceeds the numerator by $(1-p^\ai)(1-\lambda+\lambda\kappa)\geq0$, so $\underline{p}\leq1$, with $\underline{p}<1$ precisely when $p^\ai<1$. The boundary case $p^\ai=1$ gives $\underline{p}=1$ and a vacuous event $p^\doc>\underline{p}$; hence we take $p^\ai<1$. For the lower bound,
\begin{align}
p^\ai\lambda\big(\tfrac{1-\gamma}{\gamma}-\kappa\big)-(1-p^\ai)(1-\lambda+\lambda \kappa)
&=\lambda\Big[p^\ai\big(\tfrac{1}{\gamma}+\tfrac{1}{\lambda}-2\big)-\big(\tfrac{1}{\lambda}-1\big)-\kappa\Big]\\
&>\lambda\big[(1-\pi^\ai)-\kappa\big]\geq 0,
\end{align}
where the strict inequality is Assumption~\ref{assumption gamma lambda 2} for $i=\ai$ and the last uses $\kappa\leq 1-\pi^\ai$, so $\underline{p}>\tfrac12$. Hence $\underline{p}\in(\tfrac12,1)$ and $p^\doc>\underline{p}$ is attainable.

\emph{Uninterpretable AI.} The DM now observes only $\{X^\doc=(0,X^\doc(R)),C^\doc=L,D=0,A=0\}$. Marginalizing over the hidden AI signals,
\[
\Pr(A=0\mid Z=z,C,\varepsilon)=1-\pi^\ai\big[p^\ai z+(1-p^\ai)\varepsilon\big].
\]
When $X^\doc(R)=0$, this yields
\begin{align}
&\frac{\Pr(Z=1\mid X^\doc=(0,0),C^\doc=L,D=0,A=0)}{\Pr(Z=0\mid X^\doc=(0,0),C^\doc=L,D=0,A=0)}\\
&\qquad=\frac{\gamma(1-\pi^\doc)\big[(1-\lambda)(1-p^\ai\pi^\ai)+\lambda(1-\pi^\doc)(1-\pi^\ai)\big]}{(1-\gamma)\big[(1-\lambda)+\lambda(1-\pi^\doc)(1-(1-p^\ai)\pi^\ai)\big]}
\;\leq\;\frac{\gamma}{1-\gamma}<1,
\end{align}
where the weak inequality uses $p^\ai,\pi^\ai,\pi^\doc\in[0,1]$ and the strict inequality uses $\gamma<\tfrac12$. Hence $F=0$. When $X^\doc(R)=1$, the same marginalization gives
\begin{align}
&\frac{\Pr(Z=1\mid X^\doc=(0,1),C^\doc=L,D=0,A=0)}{\Pr(Z=0\mid X^\doc=(0,1),C^\doc=L,D=0,A=0)}\\
&\qquad\leq\frac{\gamma\big[p^\doc(1-\pi^\doc)\lambda+(1-p^\doc)(1-\lambda\pi^\doc)\big]}{(1-\gamma)p^\doc\lambda}<1,
\end{align}
where the weak inequality uses $1-\pi^\ai\leq 1-(1-p^\ai)\pi^\ai$ together with
\[
(1-\lambda)(1-p^\ai\pi^\ai)+\lambda(1-\pi^\doc)(1-\pi^\ai)\leq(1-\lambda\pi^\doc)(1-(1-p^\ai)\pi^\ai),
\]
which follow from $p^\ai\geq\tfrac12$ (Assumption~\ref{assumption: signals}) and $\lambda<\tfrac12$, and the strict inequality is Assumption~\ref{assumption gamma lambda 2} for $i=\doc$. Hence $F=0$ in this case as well.

Combining the two regimes, on the event $D=A=0$ and $X^\ai(C^\doc)=1$ with $p^\doc>\underline{p}$, the interpretable AI induces $F=1$ while the uninterpretable AI induces $F=0$. On any positive-probability instance of this event, because $p^\doc>\underline{p}$ is strict, the interpretable likelihood ratio strictly exceeds one: it is $+\infty$ when $X^\doc(R)=1$ and strictly above one when $X^\doc(R)=0$. Thus the interpretable posterior is strictly above $\tfrac12$. Switching from $F=0$ to $F=1$ at such a posterior strictly raises the probability of a correct decision; hence interpretability strictly raises accuracy on this event.\qed\medskip

\subsection*{Proof of Lemma~\ref{lemma follow AI or not}}
Suppose that disagreement is given by $D=0$ and $A=1$. First, consider the case in which the AI is interpretable. To identify the equilibrium in which the high-type DM makes efficient decisions, assume that this is the case. Then, according to Lemma~\ref{lemma I interpretable AI} and the fact that $p^{\doc_H}=1$, the high-type DM chooses $F=A=1$ when disagreement arises from attention differences, and $F=0\not=A$ when it arises from comprehension differences. By the off-path belief introduced in Section~\ref{sec career},
\[\Pr(\doc_H\mid F=0\not=A,\atten,\text{interpretable AI})=\Pr(\doc_H\mid F=A=1,\comp,\text{interpretable AI})=0.\] 
This implies that if the low-type DM chooses a strategy that differs from the high-type DM's, she is revealed as low-type and her reputation drops to zero. Since the low-type DM cares only about reputation, she mimics the high-type DM. Therefore,
\[\Pr(\doc_H\mid F=A=1,\atten,\text{interpretable AI})=\Pr(\doc_H\mid F=0\not=A,\comp,\text{interpretable AI})=\tau.\]
Given these reputations, the high-type DM's strategy is indeed optimal for her, confirming the equilibrium we aim to show. In equilibrium, the low-type DM follows the AI if and only if it stems from attention differences.

Second, consider the case in which the AI is uninterpretable. Again, assume that the high-type DM makes efficient decisions. According to Proposition~\ref{prop I enhance} and the fact that $p_U=1$, the high-type DM chooses $F=A=1$. This implies
\[\Pr(\doc_H\mid F=0\not=A,\text{uninterpretable AI})=0.\] 
The low-type DM, seeking to preserve her reputation, mimics the high-type DM's strategy. Therefore,
\[\Pr(\doc_H\mid F=A=1,\text{uninterpretable AI})=\tau.\]
Given these reputations, the high-type DM's strategy is indeed optimal for her, confirming the equilibrium we aim to show. In equilibrium, the low-type DM always follows the AI.
\qed\medskip

\subsection*{Proof of Proposition~\ref{prop reputation concerns}}
\begin{lemma}\label{lem reputation concerns}
    The behavior of the low-type DM differs between the interpretable and uninterpretable AI only in the following two cases: 
    \begin{enumerate}[label=(\roman*),noitemsep,topsep=0pt] 
    \item When $D=A=0$ and $X^\ai(C^\doc)=1$, the low-type DM with an interpretable AI chooses $F = 1$, whereas with an uninterpretable AI, she chooses $F = 0$.
    \item When $D=0$, $A=1$, and $X^\ai(C^\doc)=0$, the low-type DM with an interpretable AI chooses $F = 0$, whereas with an uninterpretable AI, she chooses $F = 1$.
    \end{enumerate}
\end{lemma}
\noindent\textbf{Proof.}
    When the DM cares only about reputation, the low-type DM's payoff depends only on the evaluator's posterior that she is high type. Fix any equilibrium. If the low-type DM chooses a strategy that differs from the high-type DM's, the evaluator infers that she is not high-type and assigns her zero reputation. If she instead follows the same strategy, she remains pooled with the high type and keeps a positive reputation. Hence, in equilibrium the low-type DM mimics the high-type DM. It is sufficient to characterize the high-type DM's behavior to prove the lemma. Note that we focus on the equilibrium in which the high-type DM makes efficient decisions.
    
    Observe that $\tilde{p}_U\leq \tilde{p}_I<p^{\doc_L}<p^{\doc_H}=p_U$. According to Lemma~\ref{lem II interpretable} and Proposition~\ref{prop II enhance}, when $D=1$ and $A=0$, the high-type DM never follows the AI. In contrast, Lemma~\ref{lemma I interpretable AI} and Proposition~\ref{prop I enhance} imply that when $D=0$, $A=1$, and $X^\ai(C^\doc)=1$, the high-type DM follows both the interpretable and uninterpretable AI. When $D=0$, $A=1$, and $X^\ai(C^\doc)=0$, the high-type DM only follows the uninterpretable AI. This analysis exhausts all the cases of disagreement.

    It remains to consider the cases where the DM and the AI agree on the initial decision: (1) $D = A = 0$ and (2) $D = A = 1$. Note that the high-type DM knows the true critical dimension since $p^{\doc_H} = 1$. Therefore, her final decision must depend on the observations in dimension $C^\doc$.

    In the first case, the high-type DM does not observe a positive feature in her critical dimension, i.e., $D=0	\Leftrightarrow X^\doc(C^\doc)=0$. If the AI is interpretable and $X^\ai(C^\doc)=1$, the high-type DM knows that there is a positive feature in dimension $C^\doc$. Consequently, $Z=1$, and she chooses $F=1$. However, if the AI is interpretable and $X^\ai(C^\doc)=0$, the high-type DM's belief is given by 
    $$\Pr(Z=1\mid X^\doc(C^\doc)=X^\ai(C^\doc)=0, C=C^\doc)=\frac{\gamma(1-\pi^\doc)(1-\pi^\ai)}{\gamma(1-\pi^\doc)(1-\pi^\ai)+(1-\gamma)},$$ 
    which is smaller than $\frac{1}{2}$ because $\gamma<\frac{1}{2}$. In this case, the high-type DM optimally chooses $F=0$. If the AI is uninterpretable, 
    \begin{align}
        &\Pr(Z=1\mid D=0, X^\doc(-C^\doc)=0, A=0)\leq\Pr(Z=1)=\gamma<\frac{1}{2},\text{ and }\\
        &\Pr(Z=1\mid D=0, X^\doc(-C^\doc)=1, A=0)\leq\Pr(Z=1)=\gamma<\frac{1}{2}.
    \end{align}
    In either case, the high-type DM also chooses $F=0$.

    In the second case, the high-type DM observes a positive feature in her critical dimension, i.e., $D=1\Leftrightarrow X^\doc(C^\doc)=1$. Then, $Z=1$ for sure. Hence, in this case, the high-type DM does not change her decision regardless of AI's interpretability. Summarizing this analysis completes the proof.
\qed\medskip

Following Lemma~\ref{lem reputation concerns}, we can quantify how uninterpretability changes the expected decision accuracy of the low-type DM. Recall that Lemma~\ref{lem reputation concerns} has two parts, (\rnum{1}) and (\rnum{2}). Denote the change in the low-type DM's decision accuracy as $\Delta_1$ for part (\rnum{1}) and $\Delta_2$ for part (\rnum{2}). We have
\begin{align}
    \Delta_1=&\Pr(Z=0, D=0, A=0, X^\ai(C^\doc)=1)-\Pr(Z=1, D=0, A=0, X^\ai(C^\doc)=1)\\
    =&\pi^\ai(1-\pi^\doc)[\lambda(1-\gamma)p^\ai(1-p^{\doc_L})-\gamma(1-\lambda)p^{\doc_L}(1-p^\ai)]\\
    &-\lambda\gamma(p^{\doc_L}+p^\ai-2p^{\doc_L} p^\ai)\pi^\ai(1-\pi^\doc)(1-\pi^\ai), \text{ and }\\
    \Delta_2=&\Pr(Z=1, D=0, A=1, X^\ai(C^\doc)=0)-\Pr(Z=0, D=0, A=1, X^\ai(C^\doc)=0)\\
    =&\lambda\gamma(p^{\doc_L}+p^\ai-2p^{\doc_L} p^\ai)\pi^\ai(1-\pi^\doc)(1-\pi^\ai)\\
    &+\pi^\ai[\gamma(1-\lambda)p^\ai(1-p^{\doc_L})-\lambda(1-\gamma)p^{\doc_L}(1-p^\ai)].
\end{align}
The summation, $\Delta:=\Delta_1+\Delta_2$, shares the same sign with
\begin{align}
    &p^\ai(1-p^{\doc_L})[\gamma+\lambda-2\gamma\lambda-\lambda(1-\gamma)\pi^{\doc}]-p^{\doc_L}(1-p^\ai)[\gamma+\lambda-2\gamma\lambda-\gamma(1-\lambda)\pi^{\doc}]\\
    \geq&(p^\ai-p^{\doc_L})[\gamma+\lambda-2\gamma\lambda-\gamma(1-\lambda)\pi^{\doc}]\\
    \geq&(p^\ai-p^{\doc_L})\lambda(1-\gamma)>0,
\end{align}
where the first inequality follows from $\gamma\ge\lambda$ and  $\lambda(1-\gamma)\leq\gamma(1-\lambda)$, and the second inequality follows from $\pi^{\doc}\leq 1$ and $p^\ai>p^{\doc_L}$. Let $\overline{\tau}=\frac{\Delta}{1+\Delta}$ and $\tau<\overline{\tau}$. Then, uninterpretability improves the average decision accuracy among all DMs by at least $\tau\cdot(-1)+(1-\tau)\Delta$. Since $\tau<\overline{\tau}$ and $\overline{\tau}\cdot(-1)+(1-\overline{\tau})\Delta=0$, the improvement from uninterpretability is strictly positive.
\qed\medskip

\subsection*{Consistency of Assumptions}
We take Assumptions \ref{assump:state}--\ref{assumption gamma lambda 1} as given and intend to show that Assumption \ref{assumption gamma lambda 2} and Assumption \ref{assumption career concerns} are consistent in the sense that we can find parameters that satisfy both. Let $p^{\doc_H} = 1$, and let $\pi^\doc, \pi^\ai \in (0,1)$ be arbitrary numbers such that $\pi^\doc > \pi^\ai$. Note that $p_U=1$ if $p^\ai>1-\frac{\gamma}{1-\gamma}(1-\pi^\doc)$ and Assumption~\ref{assumption gamma lambda 2} can be satisfied as long as $p^\ai$ is sufficiently close to one. Choose such a value of $p^\ai$. Since $\tilde{p}_I<p^\ai$ for any $p^\ai, \pi^\doc, \pi^\ai\in(0,1)$, we can select $p^{\doc_L}\in(\frac{1}{2},1)$ such that $\tilde{p}_I<p^{\doc_L}<p^\ai$. This construction ensures that the assumptions $p^{\doc_L} < p^\ai$, $p_U=1$, and $p^{\doc_L} > \tilde{p}_I$ are satisfied. Moreover, since $\pi^\doc > \pi^\ai$ and the AI parameters satisfy Assumption~\ref{assumption gamma lambda 2}, the same assumption is satisfied by $p^{\doc_H}$ and $\pi^\doc$. If $p^{\doc_L}$ is chosen close enough to $p^\ai$, the same assumption can be satisfied by $p^{\doc_L}$ and $\pi^\doc$ as well.
\qed\medskip

\end{APPENDIX}



\bibliographystyle{informs2014} 
\bibliography{ref_used} 

\ECSwitch

\ECHead{Online Appendices}

\phantomsection
\pdfbookmark[1]{Online Appendices}{online-appendices}
\begin{APPENDICES}

\section{AI Complementarity to the DM}\label{appendix AI complementarity}
This section formalizes the result mentioned in Remark~\ref{remark on complementarity}, which examines how the AI's attention and comprehension skills complement or substitute for the DM's comprehension skill. We define an AI skill as a \emph{complement} if it weakly increases the marginal benefit of improving the DM's comprehension on decision accuracy, and as a \emph{substitute} if it strictly decreases this marginal benefit. Our result shows that this relationship depends on the source of disagreement between the DM and the AI.
    
Suppose $\tilde{p}_I<p^\doc<p_I$, the range of the AI on which we focus in Sections~\ref{sec Main} and~\ref{sec career}. Throughout this range the AI only ever moves the DM from a negative to a positive decision, so $D\neq F$ means $D=0$ and $F=1$. Under an interpretable AI, three routes produce such a switch. Two operate under disagreement ($A=1$)---an attention difference and a comprehension difference, as in Definition~\ref{def: disagreement}. The third operates under agreement ($A=0$): by Proposition~\ref{proposition agreement}, an interpretable AI can reveal a positive feature in the DM's critical dimension even when its recommendation matches her initial decision, again inducing $F=1$. We decompose the accuracy of the DM's final decision under an interpretable AI:
\[
\begin{aligned}
\Pr(Z=F)
&= \Pr(Z=D) + \bigl[\Pr(Z=F,\,D\neq F) - \Pr(Z=D,\,D\neq F)\bigr]\\
&= \Pr(Z=D) + \bigl[\Pr(Z=1,D=0,F=1) - \Pr(Z=0,D=0,F=1)\bigr]\\
&= \Pr(Z=D) + u + v + w,
\end{aligned}
\]
where
\[
u := \Pr(Z=1,\atten) - \Pr(Z=0,\atten),\qquad
v := \Pr(Z=1,\comp) - \Pr(Z=0,\comp),
\]
\[
w := \Pr(Z=1,D=0,A=0,F=1) - \Pr(Z=0,D=0,A=0,F=1).
\]
The terms $u$ and $v$ are the AI's accuracy contributions through the two disagreement routes, and $w$ is its contribution through the agreement route of Proposition~\ref{proposition agreement}. All three are nonnegative, since along each route the DM revises to $F=1$ only when her posterior probability that $Z=1$ is at least $\tfrac{1}{2}$. The complement--substitute distinction concerns how the AI's skills interact with the DM's comprehension under disagreement, and is therefore governed by $u$ and $v$. We study how these two terms respond to an improvement in the DM's comprehension, as the following proposition characterizes.

\begin{proposition}\label{prop AI complementarity}
     Suppose $\pi^\doc,\pi^\ai\in(0,1)$. Then
\[
\frac{\partial^2 u}{\partial p^\doc\partial \pi^\ai}>0,\qquad
\frac{\partial^2 u}{\partial p^\doc\partial p^\ai}>0,\qquad
\frac{\partial^2 v}{\partial p^\doc\partial \pi^\ai}<0,\qquad
\frac{\partial^2 v}{\partial p^\doc\partial p^\ai}<0.
\]
\end{proposition}

\section{Persuasion Probability Conditional on Disagreement}
\label{Appendix conditional persuasion}

This section compares the probability that the DM is persuaded conditional on disagreement, rather than the persuasion regions compared in Proposition~\ref{prop I enhance}. Throughout, disagreement is given by $D=0$ and $A=1$.

\begin{proposition}[\textbf{Persuasion Probability Conditional on Disagreement}]
\label{prop conditional persuasion}
Suppose disagreement is given by $D=0$ and $A=1$ with $\Pr(D=0,A=1)>0$, and suppose $\pi^\doc<1$. Let $P_I(p^\doc)$ and $P_U(p^\doc)$ denote the probability that the DM is persuaded, $\Pr(F=1\mid D=0,A=1)$, under the interpretable and uninterpretable AI. Define
\[
p^\ast:=\frac{\gamma\bigl[p^\ai-\lambda(\pi^\doc+p^\ai\pi^\ai)\bigr]}
{p^\ai(\gamma+\lambda-2\pi^\ai\gamma\lambda)-\lambda(1-\pi^\ai\gamma)},
\]
whose denominator is positive under the maintained assumptions. Then $P_U(p^\doc)\geq P_I(p^\doc)$ if and only if $p^\doc\leq\max\{p_U,p^\ast\}$, with strict inequality for $p^\doc\in(p_I,p_U]$.
\end{proposition}

\section{Robustness Regarding Signal Errors}\label{Appendix signal errors}
This section shows that our results are robust to alternative assumptions about signal errors. In contrast to Assumption~\ref{assumption: signals}, which allows only Type~\Rnum{2} errors for attention signals, we generalize the analysis to settings in which attention signals may incur both Type~\Rnum{1} and Type~\Rnum{2} errors.

Denote as $\phi^i:=\Pr[X^i(j)=0\mid X(j)=0]$ the probability that $i\in\{\doc,\ai\}$ observes zero in dimension $j\in\{L,R\}$ conditional on the feature being negative, and $\xi^i:=1-\phi^i$. Thus, $\xi^i$ represents the rate of Type~\Rnum{1} error for $i$'s attention signal. We impose the following assumption that parallels with Assumption~\ref{assumption: signals}:
\begin{assumption}\label{assumption: robust}
    \begin{enumerate}[label=(\roman*), noitemsep,topsep=0pt]
        \item \label{assumption: robust 2} The DM and the AI may make Type \Rnum{2} (i.e., false negative) errors in observing features: 
        $\pi^i\coloneqq \Pr(X^i(j)=1\mid X(j)=1) \in [0, \overline{\pi}]$ 
        for $i\in\{\doc, \ai\}$ and $j\in\{L, R\}$, where $0<\overline{\pi}<1$ is a constant.
        \item \label{assumption: robust 1} The DM and the AI may make Type~\Rnum{1} errors in observing features, but the rate is sufficiently low: For $i\in\{\doc,\ai\}$,
        \begin{align}
        0\leq\xi^i<\min\bigg\{\frac{1-2\gamma}{1-\gamma},\frac{\gamma\lambda}{4}(1-\overline{\pi})^3\pi^i,
        \frac{\gamma-\lambda}{\lambda(1-\gamma)}\pi^i\bigg\}.
        \end{align}
        \item The DM and the AI may miscomprehend the critical dimension: 
        $p^i \coloneqq \Pr(C^i = C\mid C)\in[\frac{1}{2},1]$ for $i\in\{\doc, \ai\}$.
    \end{enumerate}
\end{assumption}
In the third part, this assumption coincides with Assumption~\ref{assumption: signals} regarding the accuracy of comprehension signals. In the first and the second parts, however, it differs from Assumption~\ref{assumption: signals}. Part~\ref{assumption: robust 1} allows Type~\Rnum{1} error for attention signals, as $\xi^i$ can be strictly positive. Part~\ref{assumption: robust 2} assumes that $\pi^i$, the probability that $i$ observes one conditional on the feature being positive, is bounded away from $1$. In particular, $\pi^i\leq\overline{\pi}<1$, and $\overline{\pi}$ is used in the upper bound of $\xi^i$ as seen in Part~\ref{assumption: robust 1}.

In the following analysis, we maintain Assumptions~\ref{assump:state}, \ref{assumption gamma lambda 1}, \ref{assumption gamma lambda 2}, and \ref{assumption: robust}. Our goal is to show that under these assumptions, the results in Section~\ref{sec Main} remain the same in the sense that (\rnum{1}) attention differences are still more persuasive than comprehension differences, and (\rnum{2}) uninterpretability still enhances persuasion.

\subsection{Interpretable AI}

\begin{lemma}\label{lem C1}
    The DM makes a positive initial decision (i.e., $D=1$) if and only if $X^\doc(C^\doc)=1$. The AI makes a positive recommendation (i.e., $A=1$) if and only if $X^\ai(C^\ai)=1$.
\end{lemma}

Lemma~\ref{lem C1} replicates Lemma~\ref{lem benchmark} by showing that as long as the DM or the AI does not make Type~\Rnum{1} errors too often, it still makes a positive decision/recommendation if and only if a positive feature is observed in its critical dimension.\smallskip

Next, we study AI persuasion when AI is interpretable. We adopt the same definition of the attention difference and the comprehension difference as in Section~\ref{sec Main}. 

\begin{lemma}\label{lem C2}
    Suppose that the AI is interpretable and disagreement is given by $D=0$ and $A=1$. Then, the following holds:
    \begin{enumerate}[label=(\roman*),noitemsep,topsep=0pt]
        \item When an attention difference occurs, the AI persuades the DM to change her decision from $D=0$ to $F=1$ regardless of her skill $(\pi^\doc,p^\doc)$.
        \item When a comprehension difference occurs, the AI persuades the DM only if her comprehension skill $p^\doc$ is weakly below a threshold $p'_I$.
    \end{enumerate}
\end{lemma}

Lemma~\ref{lem C2} parallels Lemma~\ref{lemma I interpretable AI}. It shows that an attention difference is persuasive regardless of the DM's skill, while a comprehension difference is persuasive only when the DM's comprehension skill is low enough. The key difference is that Lemma~\ref{lem C2} gives only a necessary condition for persuasion under a comprehension difference. The reason is that, when attention signals can equal $1$ even if the underlying feature is $0$, the AI's attention signal is no longer decisive evidence about the feature. Hence, under a comprehension difference, persuasion can depend on what the DM observes in the AI's critical dimension. In particular, the AI is more likely to persuade the DM when $X^\doc(C^\ai)=1$ than when $X^\doc(C^\ai)=0$. The threshold $p'_I$ is therefore pinned down from the case with $X^\doc(C^\ai)=1$.\smallskip

Just as Corollary~\ref{corollary 1} follows from Lemma~\ref{lemma I interpretable AI}, the following corollary follows directly from Lemma~\ref{lem C2}.

\begin{corollary}\label{corollary C1}
    The AI is more persuasive with the attention difference than the comprehension difference: $\Pr(Z=1\mid \atten, \cI^\doc)>\frac{1}{2}$ for any $(\pi^\doc,p^\doc)$, but $\Pr(Z=1\mid \comp,\cI^\doc)\geq\frac{1}{2}$ only if $p^\doc\leq p'_I$.
\end{corollary}

\subsection{Uninterpretable AI}
Finally, we study AI persuasion when AI is uninterpretable. The next proposition parallels Proposition~\ref{prop I enhance} in showing that uninterpretability can enhance persuasion.

\begin{proposition}\label{prop robust Un}
    Suppose that the AI is uninterpretable and disagreement is given by $D=0$ and $A=1$. Then, the following holds:
    \begin{enumerate}[label=(\roman*),noitemsep,topsep=0pt]
    \item There exists a threshold, $p'_U$, such that the AI persuades the DM whenever $p^\doc\leq p'_U$.
    \item $p'_U\geq p'_I$, and the inequality is strict if $\pi^\doc<1$.
    \end{enumerate}
\end{proposition}

\section{More on AI Persuasion When \texorpdfstring{$D=1$}{D=1} and \texorpdfstring{$A=0$}{A=0}}\label{Appendix: omitted}

This appendix studies the case in which the DM chooses $D=1$ but the AI recommends
$A=0$. By Lemma~\ref{lem benchmark}, the DM observes a positive feature in her critical
dimension while the AI observes a negative feature in its own. As in
Definition~\ref{def: disagreement}, three configurations are possible: (i) the AI looks
at the same dimension as the DM, and so also observes a negative feature there
($C^\ai=C^\doc$); (ii) the AI looks at a different dimension, but the DM's own attention
signal is positive there too ($C^\ai\neq C^\doc$ and $X^\doc(C^\ai)=1$, equivalently
$X^\doc=(1,1)$); and (iii) the AI looks at a different dimension, and the DM's own signal
there is negative ($C^\ai\neq C^\doc$ and $X^\doc(C^\ai)=0$). Because the DM's attention
signal carries no Type~\Rnum{1} error, in cases (i) and (ii) she already holds decisive
evidence for the positive state independent of the AI's signal; we group these as an
attention difference. Case (iii) is a comprehension difference. The classification
mirrors Definition~\ref{def: disagreement} in spirit but reverses its criterion: there,
$A=1$ confirms a positive feature in the AI's own dimension, so what mattered was the
AI's observation at the DM's dimension, $X^\ai(C^\doc)$; here, $D=1$ confirms a positive
feature in the DM's own dimension, so what matters is instead the DM's observation at
the AI's dimension, $X^\doc(C^\ai)$. The next lemma makes this precise.

\begin{lemma}\label{lem II interpretable}
    Suppose that the AI is interpretable and disagreement is given by $D=1$ and $A=0$. Then, the following hold:
    \begin{enumerate}[label=(\roman*),noitemsep,topsep=0pt]
        \item When an attention difference occurs, the AI cannot persuade the DM to change her decision from $D=1$ to $F=0$.
        \item When a comprehension difference occurs, the AI persuades the DM if and only if $X^\doc(C^\ai)=0$, and her comprehension skill, $p^\doc$, is below a threshold $\tilde{p}_I$.
    \end{enumerate}
\end{lemma}

Lemma~\ref{lem II interpretable} characterizes when an interpretable AI can persuade the DM to switch to a negative decision. If the DM initially chooses the positive decision, then she has observed a positive feature in her critical dimension. For the AI to persuade her, it must point to the other dimension, where only negative features appear. Persuasion therefore requires a comprehension difference ($C^\ai \neq C^\doc$), the absence of a positive feature in the AI's critical dimension ($X^\doc(C^\ai)=0$), and a low enough comprehension skill ($p^\doc<\tilde{p}_I$) so that the DM defers to the AI.

If the AI is uninterpretable, the DM's information set is $\tilde{\cI}^\doc=(X^\doc, C^\doc, D=1, A=0)$. She then updates as in \eqref{eq: expression} by attributing disagreement to either an attention difference or a comprehension difference. The next corollary shows that, in this setting, a comprehension difference is more persuasive than an attention difference.

\begin{corollary}\label{corollary 2}
    The AI is more persuasive when disagreement stems from a comprehension difference than an attention difference: $\Pr(Z=0\mid \comp, \tilde{\cI}^\doc)\geq \frac{1}{2}$ if $X^\doc\not=(1,1)$ and $p^\doc\leq \tilde{p}_I$, but $\Pr(Z=0\mid \atten, \tilde{\cI}^\doc)<\frac{1}{2}$.
\end{corollary}

Finally, we show how AI's uninterpretability enhances persuasion through the averaging effect and the attribution effect.

\begin{proposition}\label{prop II enhance}
    Suppose that the AI is uninterpretable and disagreement is given by $D=1$ and $A=0$. Then, if $X^\doc=(1,1)$, the AI cannot persuade the DM. If $X^\doc\neq(1,1)$, the following holds:
    \begin{enumerate}[label=(\roman*),noitemsep,topsep=0pt]
        \item  \textbf{\emph{Threshold for persuasion:}} There exists a threshold, $\tilde{p}_U$, such that if $p^\doc<\tilde{p}_U$, the AI persuades the DM. In particular, $\tilde{p}_U\leq \tilde{p}_I$.
        \item \textbf{\emph{Averaging effect:}} When $p^\doc<\tilde{p}_U$, the AI can always persuade the DM, but it might not if the AI were interpretable.
        \item \textbf{\emph{Attribution effect:}} As $\pi^\doc$ increases, $\tilde{p}_I-\tilde{p}_U$ decreases.
    \end{enumerate}
\end{proposition}

Unlike Proposition~\ref{prop I enhance}, which examines disagreement where $D=0$ and $A=1$, Proposition~\ref{prop II enhance} focuses on the case where $D=1$ and $A=0$. In this case, since an attention difference is less persuasive than a comprehension difference, the uninterpretable AI is less persuasive than the interpretable AI. However, uninterpretability still enhances persuasion. By hiding the reason for the disagreement, the uninterpretable AI can persuade the DM, even if the disagreement stems from an attention difference.

\section{Transparency Policies}
This section presents two results on transparency policies. Throughout, we
maintain Assumptions~\ref{assump:state}--\ref{assumption career concerns},
as in Section~\ref{sec career}.

\subsection{Transparency on Correct Decisions}\label{appendix Transparency on Correct Decisions}
For the next proposition, suppose that the DM's payoff is given by \[\theta\Pr(\doc_H\mid\cI_E)+(1-\theta)\bbI_{F=Z},\]
where $\cI_E$ denotes the information observed by the evaluator, and $\theta \in (0,1)$ captures the intensity of the DM's career concern. As in the main text, we continue to impose Assumptions~\ref{assump:state}--\ref{assumption career concerns}.

\begin{proposition}\label{prop transparency on correct decisions}
    \begin{enumerate}[label=(\roman*),noitemsep,topsep=0pt]
    \item When the AI is interpretable, revealing $Z$ to the evaluator weakly improves the accuracy of the final decision. 
    \item 
    For any $\tau\in(0,\overline{\tau})$, where $\overline{\tau}$ is the cutoff
in Proposition~\ref{prop reputation concerns}, there exists
$\bar\theta(\tau)<1$ such that, if $\theta>\bar\theta(\tau)$, then under the
policy in which $Z$ is revealed to the evaluator, the accuracy of the final
decision is still higher when the AI is uninterpretable than when it is interpretable.
    \end{enumerate}
\end{proposition}

\subsection{Transparency on Using AI}\label{appendix Transparency on the Influence of AI}

We now return to the pure-reputation career-concern model of
Section~\ref{sec career}. As in that section, the solution concept is Perfect
Bayesian Equilibrium, and we focus on equilibria in which the high-type DM makes
efficient decisions. In addition, when the evaluator observes only the DM's final
decision and multiple pooling equilibria exist for the low-type DM, we select
the pooling equilibrium that minimizes the low type's expected loss in decision
accuracy relative to her efficient decision. The next proposition is stated
under this equilibrium-selection convention.

\begin{proposition}\label{prop transparency on the influence of AI}
      Suppose that $\gamma=\lambda$ and the AI is interpretable. Then, the accuracy of the final decision improves if the evaluator observes only the DM's final decision, rather than every signal and decision of the DM and the AI.
\end{proposition}

\section{Effort Incentives}\label{Appendix moral hazard}

In addition to the model in Section~\ref{sec Model}, suppose the DM can, after receiving information from AI, pay a cost $c>0$ to draw an extra attention signal $X^{\mathsf{E}}$. This signal has the same structure as $X^\doc$ and, conditional on the underlying features $X(L)$ and $X(R)$, is independent of $X^\doc$ and $X^\ai$. Below we give an example in which making AI uninterpretable can improve decision accuracy by incentivizing the DM to draw  $X^{\mathsf{E}}$. For simplicity, we assume $\pi^\ai=0$. The same logic extends to $\pi^\ai>0$ as long as it is small enough.

\begin{example}\label{example effort}
    Suppose $p^\doc>p_I$ and $\pi^\ai=0$. There exist $c_1, c_2>0$ such that if $c\in(c_1,c_2)$, the following holds:
    \begin{enumerate}[label=(\roman*),noitemsep,topsep=0pt]
        \item The DM is more likely to draw $X^{\mathsf{E}}$ when AI is uninterpretable than when it is interpretable.
        \item The accuracy of the final decision is higher with uninterpretable AI than with interpretable AI.
    \end{enumerate}
\end{example}

\section{Proofs of Supplementary Results}

\subsection*{Proof of Proposition~\ref{prop AI complementarity}}

The result follows from direct calculation. Because the two dimensions are
symmetric, summing over the two values of $C$ removes the factor $1/2$.

First consider the contribution $u$ from attention differences. Decompose $u$
into the case $C^\ai=C^\doc$ and the case $C^\ai\ne C^\doc$. Let the
corresponding contributions be $u_1$ and $u_2$.

When $C^\ai=C^\doc$, an attention difference means that the AI observes a
positive feature in the DM's critical dimension while the DM misses it. Thus
\begin{align}
u_1
&:= \Pr(Z=1,\mathrm{Atten},C^\ai=C^\doc)
    - \Pr(Z=0,\mathrm{Atten},C^\ai=C^\doc)\\
&= \gamma\pi^\ai(1-\pi^\doc)
   \Big[p^\doc p^\ai
        +\lambda(1-p^\doc)(1-p^\ai)\Big]\\
&\quad
   -\lambda(1-\gamma)\pi^\ai(1-\pi^\doc)
    (1-p^\doc)(1-p^\ai)\\
&= \pi^\ai(1-\pi^\doc)
   \Big[
   \gamma p^\doc p^\ai
   -\lambda(1-2\gamma)(1-p^\doc)(1-p^\ai)
   \Big].
\end{align}
Here the first term covers $Z=1$: either the DM and the AI both correctly
comprehend the critical dimension, or they both mistakenly focus on the
non-critical dimension and $\varepsilon=1$. The second term covers $Z=0$,
which can occur only in the latter case.

When $C^\ai\ne C^\doc$, an attention difference together with $A=1$ implies
that the AI observes positive features in both dimensions. Since attention
signals have no Type I error, this requires $Z=1$ and $\varepsilon=1$. Hence
\begin{align}
u_2
&:= \Pr(Z=1,\mathrm{Atten},C^\ai\ne C^\doc)
    - \Pr(Z=0,\mathrm{Atten},C^\ai\ne C^\doc)\\
&= \gamma\lambda(\pi^\ai)^2(1-\pi^\doc)
   \Big[
   p^\doc(1-p^\ai)+(1-p^\doc)p^\ai
   \Big].
\end{align}
Therefore,
\begin{align}
u
&= \pi^\ai(1-\pi^\doc)
   \Big[
   \gamma p^\doc p^\ai
   -\lambda(1-2\gamma)(1-p^\doc)(1-p^\ai)
   \Big]\\
&\quad
   +\gamma\lambda(\pi^\ai)^2(1-\pi^\doc)
   \Big[
   p^\doc(1-p^\ai)+(1-p^\doc)p^\ai
   \Big].
\end{align}
Differentiating gives
\begin{align}
\frac{\partial^2 u}{\partial p^\doc\partial\pi^\ai}
&= (1-\pi^\doc)
   \Big[
   \gamma p^\ai+\lambda(1-2\gamma)(1-p^\ai)
   +2\gamma\lambda\pi^\ai(1-2p^\ai)
   \Big],\\
\frac{\partial^2 u}{\partial p^\doc\partial p^\ai}
&= \pi^\ai(1-\pi^\doc)
   \Big[
   \gamma-\lambda+2\gamma\lambda(1-\pi^\ai)
   \Big].
\end{align}
The first bracket is positive. Indeed, it is decreasing in $\pi^\ai$ because
$p^\ai\ge 1/2$, so it is bounded below by its value at $\pi^\ai=1$. At
$\pi^\ai=1$, it is affine in $p^\ai$ and equals
$(\gamma+\lambda-2\gamma\lambda)/2>0$ at $p^\ai=1/2$ and
$\gamma(1-2\lambda)>0$ at $p^\ai=1$. Since $\pi^\doc<1$,
$\partial^2 u/\partial p^\doc\partial\pi^\ai>0$. The second derivative is
also strictly positive because $\pi^\ai>0$, $\pi^\doc<1$, $\gamma\ge\lambda$,
and $\pi^\ai<1$.

Next consider the contribution $v$ from comprehension differences. Since
the comprehension difference forces $C^\ai\ne C^\doc$, exactly one of the DM and the AI
correctly comprehends the critical dimension. Let $v_1$ be the contribution
from $C^\doc=C$, and let $v_2$ be the contribution from $C^\ai=C$.

If $C^\doc=C$, then the AI's positive recommendation must come from the
non-critical dimension, so the event requires $\varepsilon=1$. Thus
\begin{align}
v_1
&:= \Pr(Z=1,\mathrm{Comp},C^\doc=C)
    - \Pr(Z=0,\mathrm{Comp},C^\doc=C)\\
&= \gamma\lambda p^\doc(1-p^\ai)\pi^\ai
   (1-\pi^\doc)(1-\pi^\ai)
   -\lambda(1-\gamma)p^\doc(1-p^\ai)\pi^\ai\\
&= \lambda p^\doc(1-p^\ai)\pi^\ai
   \Big[
   \gamma(1-\pi^\doc)(1-\pi^\ai)-1+\gamma
   \Big].
\end{align}
If $C^\ai=C$, then the AI's positive recommendation comes from the true
critical dimension and therefore requires $Z=1$. Hence
\begin{align}
v_2
&:= \Pr(Z=1,\mathrm{Comp},C^\ai=C)
    - \Pr(Z=0,\mathrm{Comp},C^\ai=C)\\
&= \gamma p^\ai(1-p^\doc)\pi^\ai
   \Big[
   1-\lambda+\lambda(1-\pi^\doc)(1-\pi^\ai)
   \Big].
\end{align}
Combining the two cases,
\begin{align}
v
= \pi^\ai\Big[
&\lambda p^\doc(1-p^\ai)
 \big(\gamma(1-\pi^\doc)(1-\pi^\ai)-1+\gamma\big)\\
&\quad
+\gamma p^\ai(1-p^\doc)
 \big(1-\lambda+\lambda(1-\pi^\doc)(1-\pi^\ai)\big)
\Big].
\end{align}
Differentiating gives
\begin{align}
\frac{\partial^2 v}{\partial p^\doc\partial\pi^\ai}
&=
\gamma\lambda(2p^\ai-1)(1-\pi^\doc)(2\pi^\ai-1)
-\gamma(1-\lambda)p^\ai\\
&\quad
-\lambda(1-\gamma)(1-p^\ai),\\
\frac{\partial^2 v}{\partial p^\doc\partial p^\ai}
&=
-\pi^\ai
\Big[
2\gamma\lambda(1-\pi^\doc)(1-\pi^\ai)+\gamma-\lambda
\Big].
\end{align}
For the first derivative, since $(2p^\ai-1)\ge0$ and
$(1-\pi^\doc)(2\pi^\ai-1)\le1$, we have
\[
\frac{\partial^2 v}{\partial p^\doc\partial\pi^\ai}
\le
-\gamma(1-2\lambda)p^\ai
-\lambda(1-2\gamma)(1-p^\ai)<0,
\]
where the strict inequality follows from $\lambda\le\gamma<1/2$ and
$p^\ai\ge1/2$. Finally,
$\partial^2 v/\partial p^\doc\partial p^\ai<0$ because $\pi^\ai>0$ and the
bracketed term is strictly positive under $\pi^\doc,\pi^\ai\in(0,1)$ and
$\gamma\ge\lambda$. This completes the proof. \qed\medskip

\subsection*{Proof of Proposition~\ref{prop conditional persuasion}}
Without loss of generality, suppose $C^\doc=L$; by symmetry the conditional persuasion probability is the same when $C^\doc=R$. By Lemma~\ref{lem benchmark}, $D=0$ implies $X^\doc(C^\doc)=0$, so the DM's attention signal is either $X^\doc=(0,0)$ or $X^\doc=(0,1)$. Define
\[
\begin{aligned}
B_A&:=\Pr(X^\ai(C^\doc)=1,D=0,A=1,C^\doc=L),\\
B_0&:=\Pr(X^\doc=(0,0),D=0,A=1,C^\doc=L),\\
T&:=\Pr(D=0,A=1,C^\doc=L).
\end{aligned}
\]
Since $\Pr(D=0,A=1)>0$, we have $T>0$.

\emph{Interpretable AI.} By Lemma~\ref{lemma I interpretable AI}, an attention difference persuades the DM for every $p^\doc$, whereas a comprehension difference persuades her if and only if $p^\doc\leq p_I$. Conditional on $D=0$, $A=1$, and $C^\doc=L$, the disagreement is an attention difference with probability $B_A/T$. Hence
\[
P_I(p^\doc)=\begin{dcases}1,& p^\doc\leq p_I,\\[2pt] B_A/T,& p^\doc> p_I.\end{dcases}
\]

\emph{Uninterpretable AI.} The proof of Proposition~\ref{prop I enhance} shows that the DM chooses $F=1$ for every $p^\doc$ when $X^\doc=(0,0)$, and chooses $F=1$ if and only if $p^\doc\leq p_U$ when $X^\doc=(0,1)$. Since $X^\doc=(0,0)$ occurs with probability $B_0/T$ conditional on $D=0$, $A=1$, and $C^\doc=L$,
\[
P_U(p^\doc)=\begin{dcases}1,& p^\doc\leq p_U,\\[2pt] B_0/T,& p^\doc> p_U.\end{dcases}
\]

\emph{Comparison for $p^\doc\leq p_U$.} Proposition~\ref{prop I enhance} gives $p_U\geq p_I$. If $p^\doc\leq p_I$, then $P_U=P_I=1$. If $p^\doc\in(p_I,p_U]$, then $P_U-P_I=1-B_A/T\geq 0$, and the inequality is strict whenever $B_A<T$, that is, whenever comprehension differences occur with positive probability conditional on disagreement.

\emph{Comparison for $p^\doc> p_U$.} Here $P_U-P_I=(B_0-B_A)/T$. A direct computation, marginalizing over $C$, $\varepsilon$, and the unobserved signals, gives
\[
B_0=\frac{\pi^\ai(1-\pi^\doc)}{2}\bigl[p^\ai(\gamma-\lambda)+\lambda(1-\pi^\doc\gamma)\bigr]
\]
and
\[
B_A=\frac{\pi^\ai(1-\pi^\doc)}{2}\bigl[\lambda(1-p^\ai+p^\ai\pi^\ai\gamma)+p^\doc M\bigr],
\]
where $M:=p^\ai(\gamma+\lambda-2\pi^\ai\gamma\lambda)-\lambda(1-\pi^\ai\gamma)$. Subtracting,
\[
B_0-B_A=\frac{\pi^\ai(1-\pi^\doc)}{2}\bigl[\gamma\{p^\ai-\lambda(\pi^\doc+p^\ai\pi^\ai)\}-p^\doc M\bigr].
\]
Because $p^\ai\geq\tfrac12$ and $\gamma+\lambda-2\pi^\ai\gamma\lambda>0$,
\[
M\geq\tfrac12(\gamma+\lambda-2\pi^\ai\gamma\lambda)-\lambda(1-\pi^\ai\gamma)=\frac{\gamma-\lambda}{2}\geq 0,
\]
where the last inequality is Assumption~\ref{assumption gamma lambda 1}, with equality only if $\gamma=\lambda$ and $p^\ai=\tfrac12$. At that point Assumption~\ref{assumption gamma lambda 2} for the AI requires
\[
\Bigl(\tfrac{1}{\gamma}+\tfrac{1}{\lambda}-2\Bigr)\tfrac12+\pi^\ai>\tfrac{1}{\lambda},
\]
i.e.\ $\pi^\ai>1$, which is impossible. Hence $M>0$ under the maintained assumptions. Since $\pi^\doc<1$, the prefactor $\pi^\ai(1-\pi^\doc)/2$ is strictly positive, so for $p^\doc>p_U$ the difference $B_0-B_A$ is strictly decreasing in $p^\doc$ and vanishes at $p^\ast$. Therefore $P_U(p^\doc)\geq P_I(p^\doc)$ if and only if $p^\doc\leq p^\ast$. Combined with $P_U\geq P_I$ for all $p^\doc\leq p_U$, this yields $P_U(p^\doc)\geq P_I(p^\doc)$ if and only if $p^\doc\leq\max\{p_U,p^\ast\}$.\qed\medskip

\subsection*{Proof of Lemma~\ref{lem C1}}
Consider $i\in\{\doc,\ai\}$. Without loss of generality, suppose $C^i=L$. Similar to the proof of Lemma~\ref{lem benchmark}, there are four cases to consider about $i$'s attention signal: (\rnum{1}) $X^i=(1,1)$, (\rnum{2}) $X^i=(0,0)$, (\rnum{3}) $X^i=(1,0)$, and (\rnum{4}) $X^i=(0,1)$. 

In (\rnum{1}), $\frac{\Pr(Z=1\mid X^i,C^i)}{\Pr(Z=0\mid  X^i, C^i)}=\frac{\gamma\pi^i}{(1-\gamma)\xi^i}\geq 1$, where the inequality follows from Assumption~\ref{assumption: robust}. Thus a positive decision will be made. 

In (\rnum{2}), $\frac{\Pr(Z=1\mid  X^i, C^i)}{\Pr(Z=0\mid  X^i, C^i)}=\frac{\gamma(1-\pi^i)}{(1-\gamma)(1-\xi^i)}<1$, where the inequality follows from Assumption~\ref{assumption: robust}. Thus a negative decision will be made. 

In (\rnum{3}), $\frac{\Pr(Z=1\mid  X^i, C^i)}{\Pr(Z=0\mid  X^i, C^i)}=$
\[\frac{\gamma(1-\lambda)p^i\pi^i(1-\xi^i)+\gamma(1-\pi^i)[\lambda p^i\pi^i+(1-p^i)(\lambda\pi^i+(1-\lambda)\xi^i)]}{\lambda(1-\gamma)(1-p^i)\pi^i(1-\xi^i)+(1-\gamma)\xi^i[(1-\lambda)(1-p^i)(1-\xi^i)+p^i(\lambda(1-\pi^i)+(1-\lambda)(1-\xi^i))]}.\]
Since $\gamma(1-\lambda)p^i\geq\lambda(1-\gamma)(1-p^i)$, and the second term in the denominator is strictly smaller than that in the numerator when Assumption~\ref{assumption: robust} holds, this likelihood ratio is weakly greater than one. Thus a positive decision will be made.

In (\rnum{4}), $\frac{\Pr(Z=1\mid  X^i, C^i)}{\Pr(Z=0\mid  X^i, C^i)}=$
\[\frac{\gamma\pi^i[\lambda p^i(1-\pi^i)+(1-p^i)(1-\lambda\pi^i)]+\gamma(1-\lambda)\xi^i(p^i-\pi^i)}{\lambda(1-\gamma) p^i\pi^i+(1-\gamma)\xi^i[\lambda(1-p^i)-\lambda\pi^i+(1-\lambda)(1-\xi^i)]}.\]
By Assumption~\ref{assumption gamma lambda 2}, the first term is strictly greater in the denominator than in the numerator. By Assumption~\ref{assumption: robust}, the second term is also strictly greater in the denominator. Thus the above term is strictly smaller than one, and a negative decision will be made.
\qed\medskip

\subsection*{Proof of Lemma~\ref{lem C2}}
    Without loss of generality, suppose $C^\doc=L$. According to Lemma~\ref{lem C1}, $X^\doc({L})=0$ and $X^\ai(C^\ai)=1$ when $D=0$ and $A=1$. If, in addition, $X^{\ai}({L})=1$, we get six cases:
    \begin{enumerate}[label=(\roman*),noitemsep,topsep=0pt]
        \item $X^\doc=(0,0)$, $X^\ai=(1,0)$, and $C^\ai=L$, 
        \item $X^\doc=(0,0)$, $X^\ai=(1,1)$, and $C^\ai=L$, 
        \item $X^\doc=(0,0)$, $X^\ai=(1,1)$, and $C^\ai=R$, 
        \item $X^\doc=(0,1)$, $X^\ai=(1,1)$, and $C^\ai=L$, 
        \item $X^\doc=(0,1)$, $X^\ai=(1,1)$, and $C^\ai=R$,
        \item $X^\doc=(0,1)$, $X^\ai=(1,0)$, and $C^\ai=L$.
    \end{enumerate}

    In (\rnum{1}), 
    \begin{align}
        &\frac{\Pr(Z=1\mid  X^\doc, C^\doc, X^{\ai}, C^\ai )}{\Pr(Z=0\mid  X^\doc, C^\doc, X^{\ai}, C^\ai )}\\
        =&\frac{\splitdfrac{\gamma(1-\lambda)p^\doc p^\ai(1-\pi^\doc)\pi^\ai(1-\xi^\doc)(1-\xi^\ai)}{+\gamma(1-\pi^\doc)(1-\pi^\ai)\left[\splitdfrac{\lambda p^\doc p^\ai(1-\pi^\doc)\pi^\ai}{+(1-p^\doc)(1-p^\ai)(\lambda(1-\pi^\doc)\pi^\ai+(1-\lambda)(1-\xi^\doc)\xi^\ai)}\right]}}{\splitdfrac{\lambda(1-\gamma)(1-p^\doc)(1-p^\ai)(1-\pi^\doc)\pi^\ai(1-\xi^\doc)(1-\xi^\ai)}{+(1-\gamma)(1-\xi^\doc)\xi^\ai\left[\splitdfrac{(1-\lambda)(1-p^\doc)(1-p^\ai)(1-\xi^\doc)(1-\xi^\ai)}{+p^\doc p^\ai(\lambda(1-\pi^\doc)(1-\pi^\ai)+(1-\lambda)(1-\xi^\doc)(1-\xi^\ai))}\right]}}.
    \end{align}
    This likelihood ratio is weakly greater than one because $\gamma(1-\lambda)p^\doc p^\ai\geq \lambda(1-\gamma)(1-p^\doc)(1-p^\ai)$, and the second term in the denominator is strictly smaller than that in the numerator when Assumption~\ref{assumption: robust} holds. Thus $F=1$.

    In both (\rnum{2}) and (\rnum{3}), 
    \begin{align}
        \frac{\Pr(Z=1\mid  X^\doc, C^\doc, X^{\ai}, C^\ai )}{\Pr(Z=0\mid  X^\doc, C^\doc, X^{\ai}, C^\ai )}=\frac{\gamma\pi^\ai(1-\pi^\doc)}{(1-\gamma)(1-\xi^\doc)\xi^\ai}>1,
    \end{align}
    where the strict inequality follows from Assumption~\ref{assumption: robust}. Thus $F=1$.

    In (\rnum{4}), 
    \begin{align}
        &\frac{\Pr(Z=1\mid  X^\doc, C^\doc, X^{\ai}, C^\ai )}{\Pr(Z=0\mid  X^\doc, C^\doc, X^{\ai}, C^\ai )}\\
        =&\frac{\gamma p^\doc p^\ai (1-\pi^\doc)\pi^\ai M+\gamma(1-p^\doc)(1-p^\ai)\pi^\doc\pi^\ai N}{(1-\gamma)p^\doc p^\ai (1-\xi^\doc)\xi^\ai M+(1-\gamma)(1-p^\doc)(1-p^\ai)\xi^\doc\xi^\ai N},
    \end{align}
    where $M:=\lambda\pi^\doc\pi^\ai+(1-\lambda)\xi^\doc\xi^\ai$ and $N:=\lambda(1-\pi^\doc)\pi^\ai+(1-\lambda)(1-\xi^\doc)\xi^\ai$.
    Since $\gamma(1-\pi^\doc)\pi^\ai\geq (1-\gamma)(1-\xi^\doc)\xi^\ai$ and $\gamma\pi^\doc\pi^\ai\geq(1-\gamma)\xi^\doc\xi^\ai$ when Assumption~\ref{assumption: robust} holds, the likelihood ratio is weakly greater than one. Thus $F=1$.

    In (\rnum{5}), 
    \begin{align}
        &\frac{\Pr(Z=1\mid  X^\doc, C^\doc, X^{\ai}, C^\ai )}{\Pr(Z=0\mid  X^\doc, C^\doc, X^{\ai}, C^\ai )}\\
        =&\frac{\gamma p^\doc (1-p^\ai) (1-\pi^\doc)\pi^\ai M+\gamma(1-p^\doc)p^\ai\pi^\doc\pi^\ai N}{(1-\gamma)p^\doc (1-p^\ai)(1-\xi^\doc)\xi^\ai M+(1-\gamma)(1-p^\doc)p^\ai\xi^\doc\xi^\ai N},
    \end{align}
    where $M:=\lambda\pi^\doc\pi^\ai+(1-\lambda)\xi^\doc\xi^\ai$ and $N:=\lambda(1-\pi^\doc)\pi^\ai+(1-\lambda)(1-\xi^\doc)\xi^\ai$. Then, by the same logic as in (\rnum{4}) the likelihood ratio is weakly greater than one. Thus $F=1$.

    In (\rnum{6}),
    \begin{align}
        &\frac{\Pr(Z=1\mid  X^\doc, C^\doc, X^{\ai}, C^\ai )}{\Pr(Z=0\mid  X^\doc, C^\doc, X^{\ai}, C^\ai )}\\
        =&\frac{\gamma[p^\doc p^\ai(1-\pi^\doc)\pi^\ai M+(1-p^\doc)(1-p^\ai)\pi^\doc(1-\pi^\ai)N]}{(1-\gamma)p^\doc p^\ai(1-\xi^\doc)\xi^\ai M+(1-\gamma)(1-p^\doc)(1-p^\ai)\xi^\doc(1-\xi^\ai)N},
    \end{align}
    where $M:=\lambda \pi^\doc(1-\pi^\ai)+(1-\lambda)\xi^\doc(1-\xi^\ai)$ and $N:=\lambda(1-\pi^\doc)\pi^\ai+(1-\lambda)(1-\xi^\doc)\xi^\ai$.
    Since $\gamma(1-\pi^\doc)\pi^\ai\geq (1-\gamma)(1-\xi^\doc)\xi^\ai$ and $\gamma\pi^\doc(1-\pi^\ai)\geq(1-\gamma)\xi^\doc(1-\xi^\ai)$ when Assumption~\ref{assumption: robust} holds, the likelihood ratio is weakly greater than one. Thus $F=1$. Summarizing these six cases shows that the attention difference is always persuasive.

    Then, consider the comprehension difference, i.e., $X^\ai=(0,1)$ and $C^\ai=R$. There are two cases: (\rnum{1}) $X^\doc=(0,0)$ and (\rnum{2}) $X^\doc=(0,1)$. 
    
    In (\rnum{1}),
\begin{align}\label{eq comp LR C2 p'}
    &\frac{\Pr(Z=1\mid  X^\doc, C^\doc, X^{\ai}, C^\ai )}{\Pr(Z=0\mid  X^\doc, C^\doc, X^{\ai}, C^\ai )}\\
    =&\frac{\gamma\left[\splitdfrac{p^\doc(1-p^\ai)(1-\pi^\doc)(1-\pi^\ai)(\lambda\pi^\ai(1-\pi^\doc)+(1-\lambda)\xi^\ai(1-\xi^\doc))}{+p^\ai(1-p^\doc)\pi^\ai(1-\pi^\doc)(\lambda(1-\pi^\doc)(1-\pi^\ai)+(1-\lambda)(1-\xi^\doc)(1-\xi^\ai))}\right]}{(1-\gamma)\left[\splitdfrac{p^\doc(1-p^\ai)(1-\xi^\doc)(1-\xi^\ai)(\lambda\pi^\ai(1-\pi^\doc)+(1-\lambda)(1-\xi^\doc)\xi^\ai)}{+p^\ai(1-p^\doc)(1-\xi^\doc)\xi^\ai(\lambda(1-\pi^\doc)(1-\pi^\ai)+(1-\lambda)(1-\xi^\doc)(1-\xi^\ai))}\right]},
\end{align}
which is at least one if and only if 
\begin{equation}\label{eq comp threshold C2 p'}
    p^\doc\leq p^\ai\cdot\frac{\left(\splitdfrac{\gamma\pi^\ai(1-\pi^\doc)}{-(1-\gamma)(1-\xi^\doc)\xi^\ai}\right)\left(\splitdfrac{\lambda(1-\pi^\doc)(1-\pi^\ai)}{+(1-\lambda)(1-\xi^\doc)(1-\xi^\ai)}\right)}{\left[\splitdfrac{p^\ai\left(\splitdfrac{\gamma\pi^\ai(1-\pi^\doc)}{-(1-\gamma)(1-\xi^\doc)\xi^\ai}\right)\left(\splitdfrac{\lambda(1-\pi^\doc)(1-\pi^\ai)}{+(1-\lambda)(1-\xi^\doc)(1-\xi^\ai)}\right)}{+(1-p^\ai)\left(\splitdfrac{(1-\gamma)(1-\xi^\doc)(1-\xi^\ai)}{-\gamma(1-\pi^\doc)(1-\pi^\ai)}\right)\left(\splitdfrac{\lambda\pi^\ai(1-\pi^\doc)}{+(1-\lambda)(1-\xi^\doc)\xi^\ai}\right)}\right]}.
\end{equation}

In (\rnum{2}),
\begin{align}\label{eq comp LR C2 p''}
    &\frac{\Pr(Z=1\mid  X^\doc, C^\doc, X^{\ai}, C^\ai )}{\Pr(Z=0\mid  X^\doc, C^\doc, X^{\ai}, C^\ai )}\\
    =&\frac{\gamma\left[\splitdfrac{p^\doc(1-p^\ai)(1-\pi^\doc)(1-\pi^\ai)(\lambda\pi^\doc\pi^\ai+(1-\lambda)\xi^\doc\xi^\ai)}{+p^\ai(1-p^\doc)\pi^\doc\pi^\ai(\lambda(1-\pi^\doc)(1-\pi^\ai)+(1-\lambda)(1-\xi^\doc)(1-\xi^\ai))}\right]}{(1-\gamma)\left[\splitdfrac{p^\doc(1-p^\ai)(1-\xi^\doc)(1-\xi^\ai)(\lambda\pi^\doc\pi^\ai+(1-\lambda)\xi^\doc\xi^\ai)}{+p^\ai(1-p^\doc)\xi^\doc\xi^\ai(\lambda(1-\pi^\doc)(1-\pi^\ai)+(1-\lambda)(1-\xi^\doc)(1-\xi^\ai))}\right]},
\end{align}
which is at least one if and only if 
\begin{equation}\label{eq comp threshold C2 p''}
    p^\doc\leq p^\ai\cdot\frac{\left(\splitdfrac{\gamma\pi^\doc\pi^\ai}{-(1-\gamma)\xi^\doc\xi^\ai}\right)\left(\splitdfrac{\lambda(1-\pi^\doc)(1-\pi^\ai)}{+(1-\lambda)(1-\xi^\doc)(1-\xi^\ai)}\right)}{\left[\splitdfrac{p^\ai\left(\splitdfrac{\gamma\pi^\doc\pi^\ai}{-(1-\gamma)\xi^\doc\xi^\ai}\right)\left(\splitdfrac{\lambda(1-\pi^\doc)(1-\pi^\ai)}{+(1-\lambda)(1-\xi^\doc)(1-\xi^\ai)}\right)}{+(1-p^\ai)\left(\splitdfrac{(1-\gamma)(1-\xi^\doc)(1-\xi^\ai)}{-\gamma(1-\pi^\doc)(1-\pi^\ai)}\right)\left(\splitdfrac{\lambda\pi^\doc\pi^\ai}{+(1-\lambda)\xi^\doc\xi^\ai}\right)}\right]}.
\end{equation}
Denote the right-hand side as $p'_I$. Thus in (\rnum{2}), $F=1$ if and only if $p^\doc\leq p'_I(\pi^\doc)$. A direct comparison between \eqref{eq comp threshold C2 p'} and \eqref{eq comp threshold C2 p''} reveals that $p'_I$ is weakly greater than the right-hand side of \eqref{eq comp threshold C2 p'}. As a result, given comprehension differences, the DM is persuaded only if $p^\doc\leq p'_I$.
\qed\medskip

\subsection*{Proof of Proposition~\ref{prop robust Un}}
    Without loss of generality, suppose $C^\doc=L$. There are two cases to consider about the DM's attention signal: (\rnum{1}) $X^\doc=(0,0)$ and (\rnum{2}) $X^\doc=(0,1)$. 

    In (\rnum{1}),
    \begin{align}\label{eq robust LR Un 1}
        \frac{\Pr(Z=1\mid \cI^\doc)}{\Pr(Z=0\mid \cI^\doc)}=\frac{\gamma[p^\ai\pi^\ai M+(1-p^\ai)N](1-\pi^\doc)}{(1-\gamma)[p^\ai\xi^\ai M+(1-p^\ai)N](1-\xi^\doc)},
    \end{align}
    where $M:=\lambda(1-\pi^\doc)+(1-\lambda)(1-\xi^\doc)$ and $N:=\lambda(1-\pi^\doc)\pi^\ai+(1-\lambda)(1-\xi^\doc)\xi^\ai$. It is straightforward to show that this likelihood ratio is weakly greater than one if and only if
    \begin{align}
        &(1-\pi^\doc)(1-\xi^\doc)\left[\splitdfrac{\pi^\ai(\gamma(1-\lambda)p^\ai-\lambda(1-\gamma)(1-p^\ai))}{+\xi^\ai(\gamma(1-\lambda)(1-p^\ai)-\lambda(1-\gamma)p^\ai)}\right]+\gamma\lambda(1-\pi^\doc)^2\pi^\ai\\
        \geq&(1-\gamma)(1-\lambda)(1-\xi^\doc)^2\xi^\ai.
    \end{align}
    By Assumption~\ref{assumption: robust}, the term in the square bracket is strictly positive, and the second term on the left-hand side is strictly greater than the right-hand side as $\xi^\ai<\gamma\lambda(1-\overline{\pi})^2\pi^\ai$. Thus, the above inequality holds, and the likelihood ratio \eqref{eq robust LR Un 1} is weakly greater than one. This leads to $F=1$.

    In (\rnum{2}), 
    \begin{align}
        \frac{\Pr(Z=1\mid \cI^\doc)}{\Pr(Z=0\mid \cI^\doc)}=\frac{\gamma\left[\splitdfrac{p^\doc p^\ai(1-\pi^\doc)\pi^\ai T+(1-p^\doc)(1-p^\ai)\pi^\doc Y}{+p^\doc(1-p^\ai)(1-\pi^\doc)U+p^\ai(1-p^\doc)\pi^\doc\pi^\ai O}\right]}{(1-\gamma)\left[\splitdfrac{p^\doc p^\ai(1-\xi^\doc)\xi^\ai T+(1-p^\doc)(1-p^\ai)\xi^\doc Y}{+p^\doc(1-p^\ai)(1-\xi^\doc)U+p^\ai(1-p^\doc)\xi^\doc\xi^\ai O}\right]},
    \end{align}
    where 
    \begin{align}
        T:=&\lambda\pi^\doc+(1-\lambda)\xi^\doc,\\
        Y:=&\lambda(1-\pi^\doc)\pi^\ai+(1-\lambda)(1-\xi^\doc)\xi^\ai,\\
        U:=&\lambda\pi^\doc\pi^\ai+(1-\lambda)\xi^\doc\xi^\ai,\\
        O:=&\lambda(1-\pi^\doc)+(1-\lambda)(1-\xi^\doc).
    \end{align}
    Note that this likelihood ratio is decreasing in $p^\doc$ because with respect to $p^\doc$, the partial derivative of the denominator is given by
    \[\lambda(1-\gamma)[p^\ai\xi^\ai+(1-p^\ai)\pi^\ai](\pi^\doc-\xi^\doc)>0,\]
    and the partial derivative of the numerator is given by
    \[-\gamma(1-\lambda)[p^\ai\pi^\ai+(1-p^\ai)\xi^\ai](\pi^\doc-\xi^\doc)<0,\]
    where the two inequalities follow from Assumption~\ref{assumption: robust}. Then, we can show that the likelihood ratio is at least one if and only if
    \begin{equation}\label{eq robust Un Threshold}
        p^\doc\leq\frac{p^\ai\hat{O}O+(1-p^\ai)\hat{Y}Y}{p^\ai[\hat{O}O-\hat{T}T]+(1-p^\ai)[\hat{Y}Y+\hat{U}U]},
    \end{equation}
    where
    \begin{align}
        &\hat{T}:=\gamma(1-\pi^\doc)\pi^\ai-(1-\gamma)(1-\xi^\doc)\xi^\ai,\\
        &\hat{Y}:=\gamma\pi^\doc-(1-\gamma)\xi^\doc,\\
        &\hat{U}:=(1-\gamma)(1-\xi^\doc)-\gamma(1-\pi^\doc),\\
        &\hat{O}:=\gamma\pi^\doc\pi^\ai-(1-\gamma)\xi^\doc\xi^\ai.
    \end{align}
Denote the right-hand side of \eqref{eq robust Un Threshold} as $p'_U$, and note for case (\rnum{2}) that $F=1$ if and only if $p^\doc\leq p'_U$. Then, combining cases (\rnum{1}) and (\rnum{2}), we conclude that if $p^\doc\leq p'_U$, the uninterpretable AI persuades the DM to choose $F=1$.

To show the averaging effect, suppose that $X^\doc=(0,1)$ and $p^\doc=p'_I$. 
According to Corollary~\ref{corollary C1} and the proof of Lemma~\ref{lem C2}, $\Pr(Z=1\mid  \atten, \cI^\doc)>\Pr(Z=1\mid  \comp, \cI^\doc)=\frac{1}{2}$ in this case. Equation~\eqref{eq: expression} then implies that $\Pr(Z=1\mid \cI^\doc)\geq \frac{1}{2}$, and the inequality is strict as long as $\Pr(\atten\mid \cI^\doc)\neq 0$.
However, according to the definition of $p'_U$, $\Pr(Z=1\mid \cI^\doc)=\frac{1}{2}$ when $p^\doc=p'_U$. Since $\frac{\Pr(Z=1\mid \cI^\doc)}{\Pr(Z=0\mid \cI^\doc)}$ decreases in $p^\doc$, $p'_U\geq p'_I$. Finally, note that if $\pi^\doc<1$, $\Pr(\atten\mid \cI^\doc)>0$. The same logic shows that $p'_U> p'_I$ whenever $\pi^\doc<1$. 
\qed\medskip

\subsection*{Proof of Lemma~\ref{lem II interpretable}}
According to Lemma~\ref{lem benchmark}, $X^\doc(C^\doc)=1$ and $X^\ai(C^\ai)=0$ when $D=1$ and $A=0$. If, in addition, $C^\ai =C^\doc$ or $X^\doc=(1,1)$, then the likelihood ratio of the state can be shown to be identical with \eqref{eq atten LR}. Since this likelihood ratio is weakly greater than one, the DM chooses $F=1$.

Now, suppose that $C^\ai \not=C^\doc$ and $X^\doc(C^\ai)=0$. The likelihood ratio of the state is then
\begin{equation}\label{eq comp LR-B}
    \frac{\gamma[p^\doc(1-p^\ai)(1-\lambda)+\lambda(1-\pi^\doc)(1-\pi^\ai)(p^\doc+p^\ai-2p^\doc p^\ai)]}{(1-\gamma)p^\ai(1-p^\doc)\lambda},
\end{equation}
which is smaller than one if and only if 
\begin{equation}\label{eq comp threshold-B}
    p^\doc<p^\ai\cdot\frac{(1-\gamma)/\gamma-(1-\pi^\doc)(1-\pi^\ai)}{p^\ai(1-\gamma)/\gamma+(1-p^\ai)(1-\lambda)/\lambda-(2p^\ai-1)(1-\pi^\doc)(1-\pi^\ai)}.
\end{equation}
Denote the right-hand side as $\tilde{p}_I$. We conclude that when  $C^\ai \not=C^\doc$ and $X^\doc(C^\ai)=0$, $F=0$ if and only if $p^\doc<\tilde{p}_I$.
\qed\medskip

\subsection*{Proof of Corollary~\ref{corollary 2}}
Suppose that an attention difference occurs. Notice that
\begin{equation}
\Pr(Z=0\mid \atten,\tilde{\cI}^\doc)=\Pr(Z=0\mid \cF,\atten,\tilde{\cI}^\doc)\cdot\Pr(\cF\mid \atten,\tilde{\cI}^\doc),
\end{equation}
where $\cF$ denotes the event that $(\max\{X^\doc(L),X^{\ai}(L)\},\max\{X^\doc(R),X^{\ai}(R)\})\neq(1,1)$. According to \eqref{eq atten LR}, $\Pr(Z=0\mid \cF,\atten,\tilde{\cI}^\doc)\leq\frac{1}{2}$. Since $\Pr(\cF\mid \atten,\tilde{\cI}^\doc)$ is strictly smaller than one, $\Pr(Z=0\mid \atten,\tilde{\cI}^\doc)<\frac{1}{2}$. 

Then, suppose that a comprehension difference occurs. According to \eqref{eq comp LR-B} and \eqref{eq comp threshold-B}, $\Pr(Z=0\mid \comp,\tilde{\cI}^\doc)\geq\frac{1}{2}$ if and only if $X^\doc(C^\ai)=0$ and $p^\doc\leq \tilde{p}_I$. Finally, notice that given the comprehension difference, $X^\doc(C^\ai)=0\Leftrightarrow X^\doc\neq(1,1)$. \qed\medskip

\subsection*{Proof of Proposition~\ref{prop II enhance}}
If $X^\doc=(1,1)$, it is clear that the DM never changes her decision. In the following, suppose $C^\doc=L$ and $X^\doc=(1,0)$. The likelihood ratio of the state is given by 
\begin{equation}\label{eq: likelihood ratio proof prop 2}
    \frac{\Pr(Z=1\mid \tilde{\cI}^\doc)}{\Pr(Z=0\mid \tilde{\cI}^\doc)}=\frac{\gamma[\lambda(1-\pi^\doc)(1-\pi^\ai)+p^\doc(1-p^\ai)(1-\lambda)+p^\doc p^\ai(1-\pi^\ai)(1-\lambda)]}{(1-\gamma)\lambda[p^\ai(1-p^\doc)+(1-p^\doc)(1-p^\ai)(1-\pi^\ai)]}.
\end{equation}
Because $\frac{\partial}{\partial p^\doc}[\lambda(1-\pi^\doc)(1-\pi^\ai)+p^\doc(1-p^\ai)(1-\lambda)+p^\doc p^\ai(1-\pi^\ai)(1-\lambda)]=(1-\lambda)(1-p^\ai\pi^\ai)\geq 0$, and $1-p^\doc$ is decreasing, the above likelihood ratio increases in $p^\doc$. In particular, it is strictly smaller than one if and only if 
\begin{equation}
    p^\doc<\frac{(1-\gamma)(1-\pi^\ai+p^\ai\pi^\ai)/\gamma-(1-\pi^\doc)(1-\pi^\ai)}{(1-\gamma)(1-\pi^\ai+p^\ai\pi^\ai)/\gamma+(1-\lambda)(1-p^\ai\pi^\ai)/\lambda}.
\end{equation}
Denote the right-hand side as $\tilde{p}_U$. We conclude that if $p^\doc<\tilde{p}_U$, the uninterpretable AI persuades the DM. The averaging effect then follows from this result and Lemma~\ref{lem II interpretable}.

To show $\tilde{p}_U\leq \tilde{p}_I$, suppose that $p^\doc=\tilde{p}_I$. According to Corollary~\ref{corollary 2}, $\Pr(Z=0\mid \atten, \tilde{\cI}^\doc)<\Pr(Z=0\mid \comp,\tilde{\cI}^\doc)=\frac{1}{2}$. Then,
\begin{align}
\Pr(Z=0\mid \tilde{\cI}^\doc)=&{\Pr(\comp\mid \tilde{\cI}^\doc)}\cdot\Pr(Z=0\mid  \comp, \tilde{\cI}^\doc)\\
&+\Pr(\atten\mid \tilde{\cI}^\doc)\cdot\Pr(Z=0\mid \atten, \tilde{\cI}^\doc)\leq \frac{1}{2}.
\end{align}
However, according to the definition $\tilde{p}_U$, $\Pr(Z=0\mid \tilde{\cI}^\doc)=\frac{1}{2}$ when $p^\doc=\tilde{p}_U$. Since the likelihood ratio of the state, \eqref{eq: likelihood ratio proof prop 2}, increases in $p^\doc$, $\tilde{p}_U\leq \tilde{p}_I$.

To show the attribution effect, we first notice that $\tilde{p}_U$ may be smaller than $\frac{1}{2}$. Indeed, $\tilde{p}_U\geq \frac{1}{2}$ if and only if
\begin{equation}\label{eq p4 above 0.5}
    \left[\frac{1-\lambda}{\lambda}p^\ai-\frac{1-\gamma}{\gamma}(1-p^\ai)\right]\pi^\ai\geq\frac{1}{\lambda}-\frac{1}{\gamma}+2(1-\pi^\doc)(1-\pi^\ai).
\end{equation}
We must focus on the range of $\pi^\doc$ where \eqref{eq p4 above 0.5} holds. We have
\begin{equation}
    \frac{\Pr(\comp\mid \tilde{\cI}^\doc)}{\Pr(\atten\mid \tilde{\cI}^\doc)}=\frac{\splitdfrac{\gamma(1-\lambda)p^\doc(1-p^\ai)+\lambda(1-\gamma)p^\ai(1-p^\doc)}{+\gamma\lambda(1-\pi^\doc)(1-\pi^\ai)(p^\doc+p^\ai-2p^\doc p^\ai)}}{\splitdfrac{\gamma(1-\lambda)p^\doc p^\ai(1-\pi^\ai)+\lambda(1-\gamma)(1-p^\doc)(1-p^\ai)(1-\pi^\ai)}{+\gamma\lambda(1-\pi^\doc)(1-\pi^\ai)[p^\doc p^\ai+(1-p^\doc)(1-p^\ai)]}}, \text{ and }
\end{equation}
\begin{align}
    \frac{\partial}{\partial\pi^\doc}&\frac{\Pr(\comp\mid \tilde{\cI}^\doc)}{\Pr(\atten\mid \tilde{\cI}^\doc)}=\\
    &\frac{\gamma\lambda(1-\pi^\ai)
    \left[\splitdfrac{\gamma(1-\lambda)(p^\doc)^2p^\ai(1-p^\ai)\pi^\ai+\lambda(1-\gamma)(1-p^\doc)^2p^\ai(1-p^\ai)\pi^\ai}{+p^\doc(1-p^\doc)\left[\splitfrac{\lambda(1-\gamma)(1-p^\ai)^2\pi^\ai}{+\gamma(1-\lambda)(p^\ai)^2\pi^\ai-(\gamma-\lambda)(2p^\ai-1)}\right]}\right]}
    {\left[\splitdfrac{\gamma(1-\lambda)p^\doc p^\ai(1-\pi^\ai)+\lambda(1-\gamma)(1-p^\doc)(1-p^\ai)(1-\pi^\ai)}{+\gamma\lambda(1-\pi^\doc)(1-\pi^\ai)[p^\doc p^\ai+(1-p^\doc)(1-p^\ai)]}\right]^2}.\\
\end{align}
To show this partial derivative is nonnegative, it suffices to prove
\begin{equation}\label{eq prove attribution effect for I1,0}
    \lambda(1-\gamma)(1-p^\ai)^2\pi^\ai+\gamma(1-\lambda)(p^\ai)^2\pi^\ai-(\gamma-\lambda)(2p^\ai-1)\geq 0.
\end{equation}
Note that \eqref{eq p4 above 0.5} implies the following lower bound of $\pi^\ai$:
\begin{equation}\label{eq pi_ai lower bound}
    \pi^\ai\geq\frac{\frac{1}{\lambda}-\frac{1}{\gamma}}{\left[\frac{1-\lambda}{\lambda}p^\ai-\frac{1-\gamma}{\gamma}(1-p^\ai)\right]}.
\end{equation}
By substituting \eqref{eq pi_ai lower bound} into \eqref{eq prove attribution effect for I1,0}, we can get a sufficient condition for \eqref{eq prove attribution effect for I1,0}:
\begin{equation}
    \lambda(1-\gamma)(1-p^\ai)^2+\gamma(1-\lambda)(p^\ai)^2\geq(2p^\ai-1)\left[\gamma(1-\lambda)p^\ai-\lambda(1-\gamma)(1-p^\ai)\right].
\end{equation}
It is easy to verify this inequality by collecting terms. This proves $\frac{\partial}{\partial\pi^\doc}\frac{\Pr(\comp\mid \tilde{\cI}^\doc)}{\Pr(\atten\mid \tilde{\cI}^\doc)}\geq 0$. Since $\Pr(\atten\mid \tilde{\cI}^\doc)+\Pr(\comp\mid \tilde{\cI}^\doc)=1$, the nonnegative partial derivative implies that $\Pr(\comp\mid \tilde{\cI}^\doc)$ increases in $\pi^\doc$.

Further, we calculate the following derivatives:
\begin{equation}\label{eq p3 derivative}
    \frac{\partial \tilde{p}_I}{\partial\pi^\doc}=\frac{(\frac{1-\gamma}{\gamma}+\frac{1-\lambda}{\lambda})p^\ai(1-p^\ai)(1-\pi^\ai)}{[p^\ai(1-\gamma)/\gamma+(1-p^\ai)(1-\lambda)/\lambda-(2p^\ai-1)(1-\pi^\doc)(1-\pi^\ai)]^2},
\end{equation}
\begin{equation}\label{eq p4 derivative}
    \frac{\partial \tilde{p}_U}{\partial\pi^\doc}=\frac{1-\pi^\ai}{(1-\gamma)(1-\pi^\ai+p^\ai\pi^\ai)/\gamma+(1-\lambda)(1-p^\ai\pi^\ai)/\lambda}.
\end{equation}
It is worth noting that $\frac{\partial \tilde{p}_U}{\partial\pi^\doc}\geq 0$. To show $\frac{\partial \tilde{p}_U}{\partial\pi^\doc}\geq\frac{\partial \tilde{p}_I}{\partial\pi^\doc}$, it suffices to show
\begin{align}\label{eq prove attribution effect for p4-p3}
    (\frac{1}{\gamma}+\frac{1}{\lambda}-2)&p^\ai(1-p^\ai)\left[\frac{1-\gamma}{\gamma}(1-\pi^\ai+p^\ai\pi^\ai)+\frac{1-\lambda}{\lambda}(1-p^\ai\pi^\ai)\right]\\
    &\leq\left[\frac{1-\gamma}{\gamma}p^\ai+\frac{1-\lambda}{\lambda}(1-p^\ai)-(2p^\ai-1)(1-\pi^\doc)(1-\pi^\ai)\right]^2=:K^2.
\end{align}
Denote the term in the right-hand side bracket as $K$. Note that it is increasing as a function of $\pi^\doc$ and $\pi^\ai$. And recall that by Assumption~\ref{assumption gamma lambda 2}, $\pi^\ai\geq\frac{1}{\lambda}-(\frac{1}{\gamma}+\frac{1}{\lambda}-2)p^\ai$. Then, by substituting this inequality and $\pi^\doc=0$ into $K$, we obtain one of its lower bounds:
\begin{equation}
    K\geq 2\left(\frac{1}{\gamma}+\frac{1}{\lambda}-2\right)p^\ai(1-p^\ai).
\end{equation}
In particular, $K\geq 0$. This lower bound allows us to get the following sufficient condition for \eqref{eq prove attribution effect for p4-p3}:
\begin{equation}
    2K\geq\left[\frac{1-\gamma}{\gamma}(1-\pi^\ai+p^\ai\pi^\ai)+\frac{1-\lambda}{\lambda}(1-p^\ai\pi^\ai)\right], \text{ or equivalently,}
\end{equation}
\begin{equation}\label{eq prove prove attribution effect for p4-p3 final}
    \splitdfrac{p^\ai\left[2\frac{1-\gamma}{\gamma}+\frac{1-\lambda}{\lambda}\pi^\ai-2(1-\pi^\doc)(1-\pi^\ai)\right]}{+(1-p^\ai)\left[2\frac{1-\lambda}{\lambda}+\frac{1-\gamma}{\gamma}\pi^\ai+2(1-\pi^\doc)(1-\pi^\ai)\right]}\geq\frac{1}{\gamma}+\frac{1}{\lambda}-2.
\end{equation}
The left-hand side of \eqref{eq prove prove attribution effect for p4-p3 final} can be seen as a convex combination of two terms, and each of them is greater than the right-hand side because of \eqref{eq p4 above 0.5}. This completes the proof of $\frac{\partial \tilde{p}_U}{\partial\pi^\doc}\geq\frac{\partial \tilde{p}_I}{\partial\pi^\doc}$. Therefore, $\tilde{p}_I-\tilde{p}_U$ decreases in $\pi^\doc$.
\qed\medskip

\subsection*{Comparison of the thresholds for full persuasion}
According to Proposition~\ref{prop I enhance} and Proposition~\ref{prop II enhance}, $p_I\leq p_U$ and $\tilde{p}_U\leq \tilde{p}_I$. To rank these four thresholds, we only need to compare $p_I$ and $\tilde{p}_I$. Since $\frac{1-\lambda}{\lambda}\geq\frac{1-\gamma}{\gamma}$ and $2p^\ai-1\leq 1$ in \eqref{eq comp threshold}, $p_I\geq p^\ai$. In \eqref{eq comp threshold-B}, $\tilde{p}_I\leq p^\ai$ for the same reason. This proves $\tilde{p}_I\leq p_I$. As a result, $\tilde{p}_U\leq \tilde{p}_I\leq p_I\leq p_U$. \qed\medskip

\subsection*{Proof of Proposition~\ref{prop transparency on correct decisions}}

We proceed in three steps. First, we reduce the low-type DM's problem to a
profile-by-profile mimicking decision. Second, we compare the low type's
incentive to mimic the high type when the evaluator does and does not observe
$Z$. Third, we use this comparison to prove the two parts of the proposition.

\textbf{Step 1.} Suppose first that the AI is interpretable. Let
\[
s=(X^\doc,C^\doc,D,X^\ai,C^\ai,A)
\]
denote a public signal profile before the final decision. For each type
$t\in\{H,L\}$, define
\[
\alpha_t(s):=\Pr(Z=1\mid s,\doc_t),
\qquad
f_t(s):=\bbI\{\alpha_t(s)\geq 1/2\},
\]
where $f_t(s)$ is type $t$'s efficient final decision at profile $s$.

As in Section~\ref{sec career}, we focus on equilibria in which the high-type
DM makes efficient decisions. Thus the high type chooses $F=f_H(s)$ after
profile $s$. If the low type chooses a final decision different from $f_H(s)$,
Bayes' rule assigns zero reputation to the low type, because the high type never
takes that action at profile $s$.

If $f_L(s)=f_H(s)$, the low type has no conflict between decision accuracy and
reputation: choosing $f_H(s)$ both maximizes decision accuracy and weakly
improves reputation. Hence, at such profiles, revealing $Z$ to the evaluator
does not change the low type's behavior. Therefore, the only relevant profiles
are
\[
\mathcal S:=\{s:f_L(s)\neq f_H(s)\}.
\]
This set includes the comprehension-difference histories studied in
Lemma~\ref{lem reputation concerns}, but it may also include other public
profiles. We therefore do not enumerate the profiles one by one.

Fix any $s\in\mathcal S$ with positive probability under both types. Histories
that have zero probability under the high type yield zero reputation for every
final decision and therefore create no reputational distortion. Let
\[
h:=f_H(s),
\qquad
\mu_s:=\Pr(Z=h\mid s,\doc_L),
\qquad
\eta_s:=\Pr(Z=h\mid s,\doc_H).
\]
Because the low type's efficient action differs from the high type's action,
\[
\mu_s<\frac12\leq \eta_s.
\]
Let $q$ denote the probability that the low-type DM mimics the high type and
chooses $F=h$ at profile $s$.

When the evaluator does not observe $Z$, Bayes' rule gives the reputation from
mimicking the high type as
\[
R_s^0(q)
=
\Pr(\doc_H\mid s,F=h)
=
\frac{\tau}{\tau+(1-\tau)\ell_s q},
\qquad
\ell_s:=\frac{\Pr(s\mid \doc_L)}{\Pr(s\mid \doc_H)}.
\]
If instead the low type chooses $F=1-h$, her reputation is zero.

\textbf{Step 2.} Now suppose that the evaluator also observes $Z$. Conditional
on $s$, $F=h$, and $Z=z$, Bayes' rule gives
\[
R_{s,z}^Z(q)
:=
\Pr(\doc_H\mid s,F=h,Z=z)
=
\frac{
\tau\Pr(Z=z\mid s,\doc_H)
}{
\tau\Pr(Z=z\mid s,\doc_H)
+
(1-\tau)\ell_s q\Pr(Z=z\mid s,\doc_L)
}.
\]
Thus the low type's expected reputation from mimicking the high type is
\[
R_s^Z(q)
=
\mu_s R_{s,h}^Z(q)+(1-\mu_s)R_{s,1-h}^Z(q).
\]

We now compare $R_s^Z(q)$ with $R_s^0(q)$ for a fixed $q$. By the law of total
probability,
\[
R_s^0(q)
=
\omega_s(q)R_{s,h}^Z(q)
+
(1-\omega_s(q))R_{s,1-h}^Z(q),
\]
where
\[
\omega_s(q)
=
\Pr(Z=h\mid s,F=h)
=
\frac{
\tau\eta_s+(1-\tau)\ell_s q\mu_s
}{
\tau+(1-\tau)\ell_s q
}.
\]
Since $\eta_s\geq\mu_s$,
\[
\omega_s(q)-\mu_s
=
\frac{\tau(\eta_s-\mu_s)}
{\tau+(1-\tau)\ell_s q}
\geq 0.
\]
Moreover, again because $\eta_s\geq\mu_s$,
\[
R_{s,h}^Z(q)\geq R_{s,1-h}^Z(q).
\]
Therefore,
\[
R_s^0(q)-R_s^Z(q)
=
\bigl[\omega_s(q)-\mu_s\bigr]
\bigl[R_{s,h}^Z(q)-R_{s,1-h}^Z(q)\bigr]
\geq 0.
\]
Hence, for any fixed mimicking probability $q$, revealing $Z$ weakly lowers the
low type's expected reputational payoff from mimicking the high type.

The low type's expected payoff from mimicking the high type at profile $s$ is
\[
(1-\theta)\mu_s+\theta R_s^0(q)
\]
when $Z$ is not observed, and
\[
(1-\theta)\mu_s+\theta R_s^Z(q)
\]
when $Z$ is observed. By contrast, the payoff from taking her efficient action
$1-h$ is
\[
(1-\theta)(1-\mu_s)
\]
in both cases, because choosing an action different from the high type's action
yields zero reputation.

It follows from $R_s^Z(q)\leq R_s^0(q)$ that mimicking the high type is weakly
less attractive when the evaluator observes $Z$. Let $q_s^0$ and $q_s^Z$ denote
the equilibrium mimicking probabilities at profile $s$ without and with
observability of $Z$, respectively. Since the reputational payoff from mimicking
is decreasing in $q$, the preceding comparison implies
\[
q_s^Z\leq q_s^0
\qquad\text{for every }s\in\mathcal S.
\]

Finally, conditional on $s$ and $\doc_L$, the low type's decision accuracy is
\[
q\mu_s+(1-q)(1-\mu_s)
=
1-\mu_s-q(1-2\mu_s),
\]
which is weakly decreasing in $q$ because $\mu_s<1/2$. Thus
$q_s^Z\leq q_s^0$ implies that revealing $Z$ weakly improves the low type's
decision accuracy at every profile $s\in\mathcal S$. At profiles outside
$\mathcal S$, behavior is unchanged. The high type's decision remains efficient
by construction. Therefore, when the AI is interpretable, revealing $Z$ to the
evaluator weakly improves the accuracy of the final decision. This proves
part~(\rnum{1}).

\textbf{Step 3.} We now prove part~(\rnum{2}). Fix any $\tau$ in the range
specified in the proposition. Consider the limiting case $\theta=1$, in which
the DM cares only about reputation. Then the low type strictly prefers to mimic
the high type at every public profile at which doing so gives positive
reputation, because any final decision different from the high type's decision
reveals her to be the low type and gives reputation zero. Hence, under
$\theta=1$, the low type's behavior coincides with the pure-reputation behavior
characterized in Lemma~\ref{lem reputation concerns}.

The same behavior persists for all sufficiently large $\theta<1$. To see this,
fix an AI design $j\in\{I,U\}$, where $I$ denotes an interpretable AI and $U$
denotes an uninterpretable AI. For each public profile $s$ at which the low
type's efficient action differs from the high type's action, let
$r_s^j(\tau)>0$ denote the low type's expected reputation from mimicking the high
type when all low types mimic at that profile and the evaluator observes $Z$.
Let
\[
d_s^j
:=
\Pr(Z=f_L^j(s)\mid s,\doc_L)
-
\Pr(Z=f_H^j(s)\mid s,\doc_L)
\]
denote the low type's accuracy gain from taking her efficient action rather
than mimicking the high type. If $d_s^j=0$, the profile imposes no constraint.
If $d_s^j>0$, mimicking the high type is optimal whenever
\[
\theta r_s^j(\tau)\geq (1-\theta)d_s^j,
\]
or equivalently whenever
\[
\theta
\geq
\frac{d_s^j}{d_s^j+r_s^j(\tau)}.
\]
There are only finitely many public profiles. Therefore,
\[
\bar\theta(\tau)
:=
\max_{j\in\{I,U\}}
\max_{\{s:d_s^j>0\}}
\frac{d_s^j}{d_s^j+r_s^j(\tau)}
<1.
\]
For every $\theta>\bar\theta(\tau)$, the low type mimics the high type at every
profile at which reputation and decision accuracy conflict. Hence the induced
strategies under both AI designs are the same as in the pure-reputation
equilibrium characterized by Lemma~\ref{lem reputation concerns}.

By Proposition~\ref{prop reputation concerns}, for the specified range of
$\tau$, the accuracy of the final decision is higher under the uninterpretable
AI than under the interpretable AI in that pure-reputation equilibrium. Since
the same strategies are induced for every $\theta>\bar\theta(\tau)$, the same
accuracy comparison holds here. This proves part~(\rnum{2}).
\qed\medskip

\subsection*{Proof of Proposition~\ref{prop transparency on the influence of AI}}
We proceed in two steps: (1) describe how career concerns affect the low-type DM's behavior, and (2) argue that the accuracy loss due to career concerns is greater when the evaluator observes every signal and decision of the DM and the AI than when the evaluator observes only the DM's final decision.

\textbf{Step 1:} Suppose that the AI is interpretable. If the DM cares only about decision accuracy, her behavior would follow Lemmas~\ref{lemma I interpretable AI} and \ref{lem II interpretable}. Since $p^{\doc_{H}}=1$ and $\tilde{p}_I<p^{\doc_L}<p^\ai$, the high-type DM chooses $F=1$ if and only if $\max\{X^\doc(C^\doc),X^\ai(C^\doc)\}=1$, but the low-type DM chooses $F=1$ if and only if the DM or the AI observes a positive feature in one of $C^\doc$ and $C^\ai$. This results in the following probabilities of $F=0$:
\begin{align}\label{eq interpretable and types}
    \Pr[F=0\mid\text{interpretable AI},\doc_H]&=\gamma(1-\pi^\doc)(1-\pi^\ai)+1-\gamma, \text{ and }\\
    \Pr[F=0\mid\text{interpretable AI},\doc_L]&=[p^{\doc_L}p^\ai+(1-p^{\doc_L})(1-p^\ai)][\gamma(1-\pi^\doc)(1-\pi^\ai)+1-\gamma]\\
    +&[p^{\doc_L}(1-p^\ai)+p^\ai(1-p^{\doc_L})][\gamma(1-\pi^\doc)(1-\pi^\ai)+1-\gamma]^2.
\end{align}
Here we use $\gamma=\lambda$. The probability that neither the DM nor the AI
observes a positive feature in a given dimension is
$\gamma(1-\pi^\doc)(1-\pi^\ai)+1-\gamma$ for the critical dimension, whose feature
equals $Z$, and $\lambda(1-\pi^\doc)(1-\pi^\ai)+1-\lambda$ for the non-critical
dimension, whose feature equals $\varepsilon$. These differ in general; under
$\gamma=\lambda$ they coincide. Hence the same factor
$\gamma(1-\pi^\doc)(1-\pi^\ai)+1-\gamma$ is used throughout
\eqref{eq interpretable and types}, and \eqref{eq private consulting diff},
\eqref{eq private consulting scenario d}, and
\eqref{eq private consulting scenario c} follow. We have 
\begin{align}\label{eq private consulting diff}
    &\Pr[F=0\mid\text{interpretable AI},\doc_H]-\Pr[F=0\mid\text{interpretable AI},\doc_L]\\
    =&\gamma[p^{\doc_L}(1-p^\ai)+p^\ai(1-p^{\doc_L})][\gamma(1-\pi^\doc)(1-\pi^\ai)+1-\gamma](\pi^\doc+\pi^\ai-\pi^\doc\pi^\ai)>0.
\end{align}
This shows that if both types of DMs make efficient decisions, the low-type DM would be less likely to make $F=0$.

Now consider the case in which the DM cares only about her reputation. Since we focus on the equilibrium in which the high-type DM makes efficient decisions, her behavior remains unchanged. However, the behavior of low-type DM differs. 

Suppose that the evaluator observes only the DM's final decision. Then the DM's reputation depends only on $F$, and the equilibrium must pool the high-type and low-type DMs together. In this pooling equilibrium, the evaluator's posterior belief about the DM's type must be the same after $F=0$ and $F=1$: 
\[\Pr(\doc_H\mid F=0,\text{interpretable AI})=\Pr(\doc_H\mid F=1,\text{interpretable AI}).\]
This requires the low-type DM to match the high-type DM's ex ante probability of choosing $F=0$. Since \eqref{eq private consulting diff} shows that the low-type DM would otherwise choose $F=0$ less often, she must sometimes choose $F=0$ even in scenarios where accuracy maximization calls for $F=1$. These are exactly the scenarios in which a positive feature is observed by the DM or the AI in at least one of their critical dimension. We group them as: (a) $(\max\{X^\doc(L),X^\ai(L)\},\max\{X^\doc(R),X^\ai(R)\})=(1,1)$, (b) $C^\ai=C^\doc$ in which a positive feature is observed (by the DM or the AI), (c) $C^\ai\not=C^\doc$ and a positive feature is observed only in $C^\ai$, and (d) $C^\ai\not=C^\doc$ and a positive feature is observed only in $C^\doc$. These four together exhaust the set of scenarios in which maximizing decision accuracy would require the DM to make $F = 1$. To analyze the low-type DM's behavior, we focus on the equilibrium in which she incurs the least accuracy loss (if multiple equilibria exist).

Because $p^\ai>p^{\doc_L}$ and additional positive observations raise the belief that $Z=1$, the four scenarios can be ranked as (a)$>$(b)$>$(c)$>$(d) based on the loss in decision accuracy from choosing $F = 0$. In other words, $F = 0$ hurts decision accuracy the most in scenario (a), and the least in scenario (d). 

We can compute the probability of scenario (d):
\begin{align}\label{eq private consulting scenario d}
    &\Pr[\text{scenario (d)}\mid\doc_L]\\
    =&\gamma[p^{\doc_L}(1-p^\ai)+p^\ai(1-p^{\doc_L})][\gamma(1-\pi^\doc)(1-\pi^\ai)+1-\gamma](\pi^\doc+\pi^\ai-\pi^\doc\pi^\ai).
\end{align}
By \eqref{eq private consulting diff} and \eqref{eq private consulting scenario d}, it turns out that
\begin{align}
  \Pr[F=0\mid\text{interpretable AI},\doc_H]-\Pr[F=0\mid\text{interpretable AI},\doc_L]
    =\Pr[\text{scenario (d)}\mid\doc_L].
\end{align}
This equality implies that, to mimic the high-type DM, the low-type DM just chooses $F=0$ in scenario (d), even though maximizing the decision accuracy would require $F=1$.

\textbf{Step 2:} Under the transparency regime where the evaluator observes every signal and decision of both the DM and the AI, Lemma~\ref{lem reputation concerns} implies that the low-type DM makes an inefficient choice only in scenario (c): even though accuracy maximization calls for $F=1$, she chooses $F=0$. The probability of scenario (c) is
\begin{align}\label{eq private consulting scenario c}
    &\Pr[\text{scenario (c)}\mid\doc_L]\\
    =&\gamma[p^{\doc_L}(1-p^\ai)+p^\ai(1-p^{\doc_L})][\gamma(1-\pi^\doc)(1-\pi^\ai)+1-\gamma](\pi^\doc+\pi^\ai-\pi^\doc\pi^\ai),
\end{align}
which equals \eqref{eq private consulting scenario d}. Thus, regardless of whether the evaluator observes only the DM's final decision or observes every signal and decision of both the DM and the AI, the two transparency regimes generate distortions with the same probability. 

However, the magnitude of the distortions differ. Since $p^\ai>p^{\doc_L}$, then
\[\Pr[Z=1\mid \doc_L, \text{scenario (c)}]>\Pr[Z=1\mid \doc_L, \text{scenario (d)}].\]
Using \eqref{eq private consulting scenario c}=\eqref{eq private consulting scenario d}, we obtain 
\begin{align}
    &\Pr[Z=1,\text{scenario (c)},\doc_L]-\Pr[Z=0,\text{scenario (c)},\doc_L]\\
    =&\left[\Pr[Z=1\mid\text{scenario (c)},\doc_L]-\Pr[Z=0\mid\text{scenario (c)},\doc_L]\right]\Pr[\text{scenario (c)}\mid\doc_L]\\
    >&\left[\Pr[Z=1\mid\text{scenario (d)},\doc_L]-\Pr[Z=0\mid\text{scenario (d)},\doc_L]\right]\Pr[\text{scenario (d)}\mid\doc_L]\\
    =&\Pr[Z=1,\text{scenario (d)},\doc_L]-\Pr[Z=0,\text{scenario (d)},\doc_L].
\end{align}
Therefore, when DMs have career concerns, limiting the evaluator to observe only the DM's final decision leads to a smaller accuracy loss than allowing the evaluator to observe every signal and decision of both the DM and the AI. This proves the proposition.
\qed\medskip

\subsection*{Proof of Example~\ref{example effort}}
    Because the extra attention signal never negates the observation of positive features, drawing it will not change the DM's final decision if without it, the DM would have made $F=1$. Moreover, since $p^\doc>p_I$ and $p_I\geq\tilde{p}_I$, by Lemma~\ref{lemma I interpretable AI} and Proposition~\ref{prop II enhance}, $D=1$ implies $F=1$ regardless of the interpretability of the AI. Hence, we can focus on the case when $D=0$. As  $\pi^\ai=0$, $X^{\ai}=(0,0)$ and $A=0$.

    Without loss of generality, suppose $C^\doc=L$. The original signals, $X^\doc$ and $C^\ai$, have four cases: (1) $X^\doc=(0,0)$ and $C^\ai =L$, (2) $X^\doc=(0,0)$ and $C^\ai =R$, (3) $X^\doc=(0,1)$ and $C^\ai =L$, and (4) $X^\doc=(0,1)$ and $C^\ai =R$. In each case, the DM may change her decision only if she draws another attention signal that shows $X({C^\doc})=X(L)=1$.

    For $i\in\{1,2,3,4\}$, denote the original signal profile as $\cS_i:=(X^\doc,C^\doc,X^{\ai},C^\ai )$, the extra attention signal as $X^{\mathsf{E}}$, and the expected increase in decision accuracy from this extra attention signal as $\Delta_i$. For $i=1$, we can calculate $\Delta_1$ as follows:
    \begin{equation}
        \begin{aligned}
        \Delta_1=&\Pr[Z=1, X^{\mathsf{E}}(L)=1\mid  \cS_i]-\Pr[Z=0, X^{\mathsf{E}}(L)=1\mid  \cS_i]\\
        =&\pi^\doc(1-\pi^\doc)\cdot\frac{\gamma[\mu(1-\lambda\pi^\doc)+(1-\mu)\lambda(1-\pi^\doc)]-(1-\gamma)(1-\mu)\lambda}{\gamma(1-\pi^\doc)(1-\lambda\pi^\doc)+(1-\gamma)(1-\lambda\pi^\doc)},
        \end{aligned}
    \end{equation}
    where $\mu=\frac{p^\doc p^\ai}{p^\doc p^\ai+(1-p^\doc)(1-p^\ai)}$. Similarly,
    $$\Delta_2=\pi^\doc(1-\pi^\doc)\cdot\frac{\gamma[\eta(1-\lambda\pi^\doc)+(1-\eta)\lambda(1-\pi^\doc)]-(1-\gamma)(1-\eta)\lambda}{\gamma(1-\pi^\doc)(1-\lambda\pi^\doc)+(1-\gamma)(1-\lambda\pi^\doc)},$$
    where $\eta=\frac{p^\doc(1-p^\ai)}{p^\doc(1-p^\ai)+p^\ai(1-p^\doc)}$. Notice that $\eta>\frac{1}{2}$ as $p^\ai\leq p_I<p^\doc$. Because $\mu>\eta$, $\Delta_1>\Delta_2$. In addition, because
    \begin{equation}
        \begin{aligned}
            \gamma[\eta(1-\lambda\pi^\doc)+(1-\eta)\lambda(1-\pi^\doc)]>&\gamma(1-\eta)[1+\lambda-2\lambda\pi^\doc]\quad (\eta>1-\eta)\\
            \geq&\gamma(1-\eta)(1-\lambda)\quad (\pi^\doc\leq 1)\\
            \geq&(1-\gamma)(1-\eta)\lambda,\quad (\gamma\geq\lambda)
        \end{aligned}
    \end{equation}
    we know that $\Delta_2>0$. In a similar way, we can show that $\Delta_3>\Delta_4>0$. Therefore, it is optimal for the DM to make $F=1$ whenever $X^{\mathsf{E}}({C^\doc})=X^{\mathsf{E}}(L)=1$.

    Notice the following facts about $\Delta_i$ and the DM's behavior. When AI is interpretable, the DM will draw the extra attention signal given case $(i)$ if and only if $c\leq\Delta_i$. In contrast, when AI is uninterpretable, the DM cannot distinguish between cases (1) and (2) and between cases (3) and (4). Given $X^\doc=(0,0)$, the DM will draw the extra attention signal if and only if $c\leq\Pr[\cS_1\mid  X^\doc=(0,0),C^\doc=L]\cdot\Delta_1+\Pr[\cS_2\mid  X^\doc=(0,0),C^\doc=L]\cdot\Delta_2$; and given $X^\doc=(0,1)$, she will draw the extra attention signal if and only if $c\leq\Pr[\cS_3\mid  X^\doc=(0,1),C^\doc=L]\cdot\Delta_3+\Pr[\cS_4\mid  X^\doc=(0,1),C^\doc=L]\cdot\Delta_4$.

    Now we construct $c_1$ and $c_2$, and let $c\in(c_1,c_2)$. If $\Delta_2>\Delta_4$, let $c_1=\Delta_4$ and $c_2=\min\{\Delta_2, \Pr[\cS_3\mid  X^\doc=(0,1),C^\doc=L]\cdot\Delta_3+\Pr[\cS_4\mid  X^\doc=(0,1),C^\doc=L]\cdot\Delta_4\}$. Then, when AI is uninterpretable, it is always optimal for the DM to draw the extra attention signal. In contrast, when AI is interpretable, this is optimal only in cases (1), (2), and (3).

    If $\Delta_2=\Delta_4$, let $c_1=\Delta_2$ and $c_2=\min\{\Pr[\cS_1\mid  X^\doc=(0,0),C^\doc=L]\cdot\Delta_1+\Pr[\cS_2\mid  X^\doc=(0,0),C^\doc=L]\cdot\Delta_2, \Pr[\cS_3\mid  X^\doc=(0,1),C^\doc=L]\cdot\Delta_3+\Pr[\cS_4\mid  X^\doc=(0,1),C^\doc=L]\cdot\Delta_4\}$. Then, when AI is uninterpretable, it is always optimal for the DM to draw the extra attention signal. In contrast, when AI is interpretable, this is optimal only in cases (1) and (3).

    If $\Delta_4>\Delta_2$, let $c_1=\Delta_2$ and $c_2=\min\{\Pr[\cS_1\mid  X^\doc=(0,0),C^\doc=L]\cdot\Delta_1+\Pr[\cS_2\mid  X^\doc=(0,0),C^\doc=L]\cdot\Delta_2,\Delta_4\}$. Then, when AI is uninterpretable, it is always optimal for the DM to draw the extra attention signal. In contrast, when AI is interpretable, this is optimal only in cases (1), (3), and (4).

    Finally, to show that the DM's decision accuracy is higher when the AI is uninterpretable than interpretable, recall that $p^\doc>p_I$. By Lemma~\ref{lemma I interpretable AI} and Proposition~\ref{prop I enhance}, if the DM does not draw $X^{\mathsf{E}}$, the DM's final decision remains the same as her initial decision regardless of the interpretability of the AI. Then, recall that the construction in the previous three paragraphs shows that as $\Delta_i>0$ for any $i$, drawing $X^{\mathsf{E}}$ improves decision accuracy in all the four cases, and the DM draws $X^{\mathsf{E}}$ in more cases when AI is uninterpretable than when it is interpretable. Hence, we can conclude that the accuracy of the DM's final decision is higher with uninterpretable AI than with interpretable AI.
\qed

\end{APPENDICES}

\end{document}